\documentclass[preprint2]{aastex}%

\usepackage{amsmath}
\makeindex

\makeatletter % This code stolen from ltxguide.cls:

\let\o@verbatim\verbatim
\def\verbatim{%
  \ifhmode\unskip\par\fi
  \ifx\@currsize\normalsize
     \small
  \fi
  \o@verbatim
}

\renewcommand \verbatim@font {%
  \normalfont \ttfamily
  \catcode`\<=\active
  \catcode`\>=\active
}

\RequirePackage{shortvrb}
\MakeShortVerb{\|}

\begingroup
  \catcode`\<=\active
  \catcode`\>=\active
  \gdef<{\@ifnextchar<\@lt\@meta}
  \gdef>{\@ifnextchar>\@gt\@gtr@err}
  \gdef\@meta#1>{\m{#1}}
  \gdef\@lt<{\char`\<}
  \gdef\@gt>{\char`\>}
\endgroup
\def\@gtr@err{%
   \ClassError{ltxguide}{%
      Isolated \protect>%
   }{%
      In this document class, \protect<...\protect>
      is used to indicate a parameter.\MessageBreak
      I've just found a \protect> on its own.
      Perhaps you meant to type \protect>\protect>?
   }%
}
\def\verbatim@nolig@list{\do\`\do\,\do\'\do\-}

\newcommand{\m}[1]{\mbox{$\langle$\it #1\/$\rangle$}}

\def\cmd#1{\cs{\expandafter\cmd@to@cs\string#1}}
\def\cmd@to@cs#1#2{\char\number`#2\relax}
\DeclareRobustCommand\cs[1]{\texttt{\char`\\#1}}
\def\GetFileInfo#1{%
  \def\filename{#1}%
  \def\@tempb##1 ##2 ##3\relax##4\relax{%
    \def\filedate{##1}%
    \def\fileversion{##2}%
    \def\fileinfo{##3}}%
  \edef\@tempa{\csname ver@#1\endcsname}%
  \expandafter\@tempb\@tempa\relax? ? \relax\relax}

\def\mybf{}

\makeatother
\title{%
 Probing the Masses and Radii of Donor Stars in Eclipsing X-ray Binaries with the Swift Burst Alert Telescope
}%
\author{Joel~B.~Coley\altaffilmark{1,2},
Robin~H.~D.~Corbet\altaffilmark{1,2},
Hans~A.~Krimm\altaffilmark{3,4}}
\email{jcoley1@umbc.edu}
\altaffiltext{1}{University of Maryland,Baltimore County, MD, USA}
\altaffiltext{2}{CRESST/Mail Code 662, X-ray Astrophysics Laboratory, NASA Goddard Space Flight Center, Greenbelt, MD 20771, USA}
\altaffiltext{3}{Universities Space Research Association, 7178 Columbia Gateway Drive, Columbia, MD 21046, USA}
\altaffiltext{4}{CRESST/Mail Code 661, X-ray Astroparticle Physics Laboratory, NASA Goddard Space Flight Center, Greenbelt, MD 20771, USA}

\expandafter\GetFileInfo\expandafter{\jobname.tex}%

\begin{document}

\begin{abstract}

Physical parameters of both the mass donor and compact object can be constrained in X-ray binaries with well-defined eclipses, as our survey of wind-fed supergiant X-ray binaries (SGXBs) IGR J16393-4643, IGR J16418-4532, IGR J16479-4514, IGR J18027-2016 and XTE J1855-026 reveals.  Using the orbital period and Kepler's third law, we express the eclipse half-angle in terms of radius, inclination angle and the sum of the masses.  Pulse-timing and radial velocity curves can give masses of both the donor and compact object as in the case of the ``double-lined" {\mybf binaries IGR J18027-2016 and XTE J1855-026}.  The eclipse half angles are {\mybf 15$^{+3}_{-2}$}, {\mybf 31.7$^{+0.7}_{-0.8}$}, {\mybf 32$\pm$2}, {\mybf 34$\pm$2} and {\mybf 33.6$\pm$0.7} degrees for IGR J16393-4643, IGR J16418-4532, IGR J16479-4514, IGR J18027-2016 and XTE 1855-026, respectively.  In wind-fed systems, the primary not exceeding the Roche-lobe size provides an upper limit on system parameters.  In IGR J16393-4643, spectral types of B0 V or B0-5 III are found to be consistent with the eclipse duration and Roche-lobe, but the previously proposed donor stars in IGR J16418-4532 and IGR J16479-4514 were found to be inconsistent with the Roche-lobe size.  Stars with spectral types O7.5 I and earlier are possible.  For IGR J18027-2016, the mass and radius of the donor star lie between {\mybf 18.6--19.4}\,$M_\sun$ and {\mybf 17.4--19.5}\,$R_\sun$.  We constrain the neutron star mass between 1.37--{\mybf 1.43}\,$M_\sun$.  {\mybf We find the mass and radius of the donor star in XTE J1855-026 to lie between 19.6--20.2\,$M_\sun$ and 21.5--23.0\,$R_\sun$.  The neutron star mass was constrained to 1.77--1.82\,$M_\sun$.}  Eclipse profiles are asymmetric in IGR J18027-2016 and XTE J1855-026, which we attribute to {\mybf accretion wakes}.

\end{abstract}

\section{Introduction}

High-Mass X-ray Binaries (HMXBs) are relatively young systems, which consist of a compact object (neutron star or black hole) and an early-type OB star orbiting the common center of mass.  First discovered in the 1970s, HMXBs are split into two individual classes--Be X-ray binaries (BeXBs) and supergiant X-ray binaries (SGXBs).  BeXBs are transient systems where the compact object is in a wide, typically eccentric, orbit (P$\gtrsim$10\,days) around a rapidly rotating non-supergiant B-type star.  The primary mode of mass transfer occurs when the compact object passes through the circumstellar decretion disc of the mass donor.  In SGXBs, the compact object is in a short ($\sim$1--42\,day) orbit around an OB-supergiant where the accretion mechanism is either via the powerful stellar wind and/or Roche-lobe overflow.  While the eccentricity in most SGXBs with short orbital periods is near zero, some SGXBs host compact objects in highly eccentric orbits \citep[e.g. GX 301-2][]{2014MNRAS.441.2539I}.  In wind-accretors, the X-ray luminosity is on the order of 10$^{35}$--10$^{36}$\,erg s$^{-1}$.  However, a much higher luminosity is found in systems where the donor fills its Roche-lobe, $\sim$10$^{38}$\,erg s$^{-1}$ \citep{2004RMxAC..21..128K}.

Many HMXBs host a neutron star where modulation is often seen at its rotation period.  The mass of the neutron star can be constrained in eclipsing X-ray pulsars, which can lead to an improved understanding of the neutron star Equation of State \citep[][and references therein]{2011A&A...532A.124M}.  Currently, the neutron star mass has been constrained in 10 XRBs where the lower limit for each object corresponds to edge-on orbital inclinations and the upper limit is calculated when the donor star just fills its Roche-lobe.  While over 100 Equations of State have been proposed \citep{2006Msngr.126...27K}, only one model is physically correct \citep{2010A&A...509A..79M,2011A&A...532A.124M}.

The mass ratio between the neutron star and donor star is equal to the ratio between the semi-amplitude of the radial velocities of the {\mybf donor star, $K_{\rm O}$, and the neutron star, $K_{\rm X}$}.  Throughout the paper, we use a definition of mass ratio, $q$, as the ratio of the compact object mass to that of the donor star mass {\mybf \citep{1984ARA&A..22..537J}}.  The orbital period of the binary, $P$, and $K_{\rm X}$ can be calculated using pulse-timing analysis \citep[][and references therein]{2005A&A...441..685V}, {\mybf which is analogous to measuring the Doppler shift of spectral lines in the optical and/or near-infrared \citep{1984ARA&A..22..537J}.}  The projected semi-major axis can be determined from the semi-amplitude of the radial velocity of the neutron star.  The semi-amplitude of the radial velocity of the donor star can be determined using optical {\mybf and/or near-infrared} spectroscopic information.

{\mybf Eclipse measurements can also be exploited as timing markers to determine the binary orbital evolution of HMXBs.  A significant orbital period derivative, $\dot{P}$, was previously found in several eclipsing HMXBs \citep[e.g 4U 1700-377; SMC X-1; Cen X-3; LMC X-4 and OAO 1657-415,][]{1996ApJ...459..259R,2010MNRAS.401.1532R,2015A&A...577A.130F} and can be used to investigate the orbital evolution over long periods of time.  While several contending theories to explain the orbital decay have been investigated, the most probable explanations involve tidal interaction and rapid mass transfer between the components of the binary systems \citep[][and references therein]{2015A&A...577A.130F}.}

The \textsl{Swift} Burst Alert Telescope (BAT), sensitive to X-rays in the 15--150\,keV band \citep{2005SSRv..120..143B}, provides an excellent way to study highly absorbed SGXBs.  The large absorption found in these systems is problematic for instruments such as the \textsl{Rossi X-ray Timing Explorer (RXTE)} All Sky Monitor (ASM), which operated in the 1.5--12\,keV band \citep{2011ApJS..196....6L}.  The sensitivity to higher energy X-rays allows \textsl{Swift}-BAT to peer through this absorption \citep[][and references therein]{2013ApJ...778...45C}.

We present here constraints on the mass and radius of the donor star in eclipsing XRBs. The probability of an eclipse in a supergiant XRB with an orbital period less than 20\,d can be expressed in terms of the orbital period, mass of the donor star, and radius of the donor star \citep[Equation 1;][]{2002ApJ...581.1293R}.  The probability of an eclipse in long-period XRBs is low.  Using a literature search, {\mybf we determined} HXMB systems where BAT observations can {\mybf significantly} improve the {\mybf the properties of the stellar components and orbital evolution of the systems using eclipsing properties. Five} eclipsing XRBs were identified: IGR J16393-4643, IGR J16418-4532, IGR J16479-4514, IGR J18027-2016 and XTE J1855-026, {\mybf which are all highly obscured SGXBs}.  {\mybf We note while} the masses of both the donor and compact object had previously been constrained in IGR J18027-2016, the error on the eclipse half-angle was large at 4.5$\degr$ \citep{2005A&A...439..255H}.  The error {\mybf estimates concerning the radius and mass of the donor star are} significantly improved in our analysis.

This paper is structured in the following order:  \textsl{Swift} BAT observations and the description of the eclipse model are presented in Section 2; Section 3 focuses on individual sources that are known to be eclipsing.  Section 4 presents a discussion of the results and the conclusions are outlined in Section 5.  If not stated otherwise, the uncertainties and limits presented in the paper are at the 1$\sigma$ confidence level.

\section{Data Analysis and Modelling}

\subsection{Swift BAT}

The BAT on board the \textsl{Swift} spacecraft is a hard X-ray telescope operating in the 15--150\,keV energy band \citep{2005SSRv..120..143B}.  The detector is composed of CdZnTe where the detecting area and field of view (FOV) are 5240\,cm$^2$ and 1.4\,sr (half-coded), respectively \citep{2005SSRv..120..143B}.  The BAT provides an all-sky hard X-ray survey with a sensitivity of $\sim$1\,mCrab \citep{2010ApJS..186..378T}.  The Crab produces $\sim$0.045\,counts cm$^{-2}$ s$^{-1}$ over the entire energy band.

We analyzed BAT data obtained during the time period MJD\,53416--56745 (2005 February 15--2014 March 29).  Light curves were retrieved using the extraction of the BAT transient monitor data available on the NASA GSFC HEASARC website\footnote{http://heasarc.gsfc.nasa.gov/docs/swift/results/transients/} \citep{2013ApJS..209...14K}, which includes orbital and daily-averaged light curves.  We used the orbital light curves in the 15--50\,keV energy band in our analysis, {\mybf which have typical exposures of $\sim$6\,min} (see Section~\ref{Five Eclipsing HMXBs}).  {\mybf The short exposures, which are somewhat less than typical \textsl{Swift} pointing times ($\sim$20\,min), can arise due to the observing plan of \textsl{Swift} itself where BAT is primarily tasked to observe gamma-ray bursts \citep{2013ApJS..209...14K}.}

The light curves were further screened to exclude bad quality points.  We only considered data where the data quality flag (``DATA\_FLAG") was set to 0.  Data flagged as ``good" are sometimes suspect, where a small number of data points with very low fluxes and implausibly small uncertainties were found \citep{2013ApJ...778...45C}.  These points were removed from the light curves.  {\mybf We corrected the photon arrival times to the solar system barycenter.  We used the scripts made available on the Ohio State Astronomy webpage\footnote{http://astroutils.astronomy.ohio-state.edu/time/.}  In this paper, the barycenter-corrected times are referred to as Barycenter Modified Julian Date (BMJD).}

We initially derived the orbital period for each XRB in our sample using Discrete Fourier transforms (DFTs) of the light curves to search for periodicities in the data.  We weighted the contribution of each data point to the power spectrum by its uncertainty, using the ``semi-weighting" technique \citep{2007PThPS.169..200C, 2013ApJ...778...45C}, where the error bars on each data point and the excess variability of the light curve are taken into account.  We derived uncertainties on the orbital periods using the expression given in \citet{1986ApJ...302..757H}.

\subsection{Eclipse Modeling}
\label{Eclipse Modeling}

{\mybf We initially modeled the eclipses using a symmetric ``step and ramp" function (see Table~\ref{Step and Ramp Model}) where the intensities are assumed to remain constant before ingress, during eclipse and after egress and follow a linear trend during the ingress and egress transitions \citep{2014ApJ...793...77C}. The count rates before ingress, during eclipse and after egress were each considered to be free parameters and were fit as follows: $C_{\rm ing}$ was fit from binary phase $\phi=$-0.2 to the start of ingress, $C_{\rm ecl}$ was fit during eclipse and $C_{\rm eg}$ was fit from the end of egress to phase $\phi=$0.2 (see Equation~\ref{Flux Equation}).  While we find the eclipse profiles of IGR J16393-4643, IGR J16418-4532 and IGR J16479-4514 to be symmetric within errors, the profiles show some asymmetry in the cases of IGR J18027-2016 and XTE J1855-026.  We note that a symmetric ``step and ramp" function could lead to systematic errors and we therefore fit the eclipse profiles using an asymmetric ``step and ramp" function (see Equantion~\ref{Flux Equation} and Table~\ref{Asymmetric Step and Ramp Model}).  The parameters in this model are as follows: the phases corresponding to the start of ingress and start of egress, $\phi_{\rm ing}$ and $\phi_{\rm eg}$, the duration of ingress, $\Delta{\phi}_{\rm ing}$, the duration of egress, $\Delta{\phi}_{\rm eg}$, the pre-ingress count rate, $C_{\rm ing}$, the post-egress count rate, $C_{\rm eg}$, and the count rate during eclipse, $C_{\rm ecl}$.  The eclipse duration and mid-eclipse phase are calculated using Equations~\ref{Half Angle Equation} and~\ref{Mid Eclipse Equation}, respectively.}

\begin{equation}
\label{Flux Equation}
  C =
  \left\{
    \begin{array}{ll}
      C_{\rm ing}, & -0.2 \leq \phi \leq \phi_{\rm ing} \\
      (\frac{C_{\rm ecl}-C_{\rm ing}}{\Delta{\phi}_{\rm ing}}) (\phi-\phi_{\rm ing})+C_{\rm ing}, & \phi_{\rm ing} \leq \phi \leq \phi_{\rm ing}+\Delta{\phi}_{\rm ing} \\
      C_{\rm ecl}, & \phi_{\rm ing}+\Delta{\phi}_{\rm ing} \leq \phi \leq \phi_{\rm egr} \\
      (\frac{C_{\rm egr}-C_{\rm ecl}}{\Delta{\phi}_{\rm eg}}) (\phi-\phi_{\rm egr})+C_{\rm ecl}, & \phi_{\rm egr} \leq \phi \leq \phi_{\rm egr}+\Delta{\phi}_{\rm eg} \\
      C_{\rm eg}, & \phi_{\rm egr}+\Delta{\phi}_{\rm eg} \leq \phi \leq 0.2 \\
    \end{array}
  \right.
\end{equation}

\begin{equation}
\label{Half Angle Equation}
\Delta{\phi}_{\rm ecl}=\phi_{\rm egr}-(\phi_{\rm ing}+\Delta{\phi}_{\rm ing})
\end{equation}

\begin{equation}
\label{Mid Eclipse Equation}
\phi_{\rm mid}=\frac{1}{2}(\phi_{\rm egr}+(\phi_{\rm ing}+\Delta{\phi}_{\rm ing}))
\end{equation}

% Table 1.  This is the Eclipse Model Parameters
\begin{deluxetable}{cccccc}
\tablecolumns{6}
\tabletypesize{\small}
\tablewidth{0pc}
\tablecaption{Eclipse Model Parameters, Assuming a Symmetric Eclipse Profile}
\tablehead{
\colhead{Model Parameter} & \colhead{IGR J16393-4643} & \colhead{IGR J16418-4532} & \colhead{IGR J16479-4514} & \colhead{IGR J18027-2016} & \colhead{XTE J1855-026}}
\startdata
$\phi_{\rm ing}$ & {\mybf -0.079$_{-0.014}^{+0.006}$} & {\mybf -0.107$\pm$0.002} & {\mybf -0.114$_{-0.004}^{+0.003}$} & {\mybf -0.150$_{-0.003}^{+0.005}$} & {\mybf -0.131$_{-0.002}^{+0.001}$} \\
$\Delta{\phi}$ & {\mybf 0.040$_{-0.008}^{+0.009}$} & {\mybf 0.019$_{-0.003}^{+0.002}$} & {\mybf 0.029$_{-0.004}^{+0.003}$} & {\mybf 0.053$_{-0.003}^{+0.004}$} & {\mybf 0.038$\pm$0.002} \\
$C$$^a$ & {\mybf 1.27$\pm$0.04} & {\mybf 1.18$\pm$0.05} & {\mybf 1.05$\pm$0.06} & {\mybf 1.59$\pm$0.06} & {\mybf 2.64$\pm$0.06} \\
$\phi_{\rm egr}$ & {\mybf 0.048$_{-0.005}^{+0.004}$} & {\mybf 0.088$_{-0.001}^{+0.002}$} & {\mybf 0.092$_{-0.004}^{+0.003}$} & {\mybf 0.098$_{-0.004}^{+0.002}$} & {\mybf 0.094$\pm$0.002} \\
$C_{\rm ecl}$$^a$ & {\mybf 0.71$\pm$0.06} & {\mybf 0.00$\pm$0.05} & {\mybf 0.03$\pm$0.05} & {\mybf 0.17$\pm$0.04} & {\mybf -0.04$\pm$0.05} \\
\tableline
$\Delta{\phi}_{\rm ecl}$ & {\mybf 0.09$^{+0.01}_{-0.02}$} & {\mybf 0.175$\pm$0.003} & {\mybf 0.177$^{+0.005}_{-0.007}$} & {\mybf 0.196$^{+0.007}_{-0.005}$} & {\mybf 0.187$\pm$0.003} \\
$P_{\rm orb}^b$ & {\mybf 4.23794$\pm$0.00007} & {\mybf 3.73880$\pm$0.00002} & {\mybf 3.31965$\pm$0.00006} & {\mybf 4.56999$\pm$0.00005} & {\mybf 6.07410$\pm$0.00004} \\
$\dot{P}_{\rm orb}^c$ & {\mybf -5$\pm$4} & {\mybf 0.7$\pm$1.0} & {\mybf 3$\pm$2} & {\mybf -2$\pm$2} & {\mybf 0.5$\pm$1.0} \\
$T_{\rm mid}^d$ & {\mybf 55074.99$_{-0.04}^{+0.02}$} & {\mybf 55087.714$\pm$0.006} & {\mybf 55081.571$^{+0.009}_{-0.012}$} & {\mybf 55083.79$_{-0.01}^{+0.02}$} & {\mybf 55079.055$_{-0.009}^{+0.010}$} \\
$\Theta_{\rm e}^e$ & {\mybf 16$^{+2}_{-3}$} & {\mybf 31.5$\pm$0.6} & {\mybf 31.9$^{+0.9}_{-1.3}$} & {\mybf 35$\pm$1} & {\mybf 33.6$^{+0.6}_{-0.5}$} \\
\tableline
$\chi^2_\nu$ (dof) & {\mybf 1.13(77)} & {\mybf 0.84(77)} & {\mybf 1.03(77)} & {\mybf 0.93(77)} &  {\mybf 1.23(77)} \\
\enddata
\tablecomments{\\*
$^a$ Units are 10$^{-3}$\,counts cm$^{-2}$ s$^{-1}$. \\*
$^b$ Refined orbital periods using an $O-C$ analysis.  Units are days. \\*
$^c$ The orbital period derivative at the 90$\%$ confidence interval found using an $O-C$ analysis.  Units are 10$^{-7}$\,d d$^{-1}$. \\*
$^d$ Units are barycentered Modified Julian Time (BMJD). Phase 0 is defined as eclipse center. \\*
$^e$ Units are degrees.}
\label{Step and Ramp Model}
\end{deluxetable}

% Table 2.  This is the Eclipse Model Parameters
\begin{deluxetable}{cccccc}
\tablecolumns{6}
\tabletypesize{\small}
\tablewidth{0pc}
\tablecaption{{\mybf Eclipse Model Parameters, Assuming an Asymmetric Eclipse Profile}}
\tablehead{
\colhead{{\mybf Model Parameter}} & \colhead{{\mybf IGR J16393-4643}} & \colhead{{\mybf IGR J16418-4532}} & \colhead{{\mybf IGR J16479-4514}} & \colhead{{\mybf IGR J18027-2016}} & \colhead{{\mybf XTE J1855-026}}}
\startdata
{\mybf $\phi_{\rm ing}$} & {\mybf -0.092$_{-0.005}^{+0.008}$} & {\mybf -0.108$\pm$0.002} & {\mybf -0.118$\pm$0.003} & {\mybf -0.160$\pm$0.005} & {\mybf -0.136$\pm$0.002} \\
{\mybf $\Delta{\phi}_{\rm ing}$} & {\mybf 0.06$\pm$0.01} & {\mybf 0.020$\pm$0.003} & {\mybf 0.024$\pm$0.008} & {\mybf 0.08$\pm$0.01} & {\mybf 0.042$\pm$0.003} \\
{\mybf $\Delta{\phi}_{\rm eg}$} & {\mybf 0.039$\pm$0.006} & {\mybf 0.014$_{-0.004}^{+0.003}$} & {\mybf 0.026$_{-0.003}^{+0.007}$} & {\mybf 0.018$_{-0.004}^{+0.005}$} & {\mybf 0.038$\pm$0.003} \\
{\mybf $C_{\rm ing}$$^a$} & {\mybf 1.28$\pm$0.06} & {\mybf 1.07$_{-0.06}^{+0.07}$} & {\mybf 1.01$\pm$0.08} & {\mybf 1.43$\pm$0.08} & {\mybf 2.44$\pm$0.08} \\
{\mybf $C_{\rm eg}$$^a$} & {\mybf 1.26$\pm$0.06} & {\mybf 1.22$\pm$0.07} & {\mybf 1.04$\pm$0.08} & {\mybf 1.67$\pm$0.08} & {\mybf 2.79$_{-0.08}^{+0.09}$} \\
{\mybf $\phi_{\rm egr}$} & {\mybf 0.049$_{-0.006}^{+0.004}$} & {\mybf 0.089$\pm$0.002} & {\mybf 0.086$\pm$0.003} & {\mybf 0.112$\pm$0.002} & {\mybf 0.092$_{-0.002}^{+0.001}$} \\
{\mybf $C_{\rm ecl}$$^a$} & {\mybf 0.69$\pm$0.06} & {\mybf 0.00$\pm$0.05} & {\mybf 0.05$\pm$0.05} & {\mybf 0.14$\pm$0.05} & {\mybf -0.05$\pm$0.05} \\
\tableline
{\mybf $\Delta{\phi}_{\rm ecl}$} & {\mybf 0.09$^{+0.02}_{-0.01}$} & {\mybf 0.176$\pm$0.004} & {\mybf 0.180$\pm$0.009} & {\mybf 0.19$\pm$0.01} & {\mybf 0.186$\pm$0.004} \\
{\mybf $P_{\rm orb}^b$} & {\mybf 4.23810$\pm$0.00007} & {\mybf 3.73881$\pm$0.00002} & {\mybf 3.31961$\pm$0.00004} & {\mybf 4.56988$\pm$0.00006} & {\mybf 6.07413$\pm$0.00004} \\
{\mybf $\dot{P}_{\rm orb}^c$} & {\mybf -6$\pm$4} & {\mybf 0.2$\pm$0.8} & {\mybf 3$\pm$2} & {\mybf -3$\pm$4} & {\mybf 0.5$\pm$1.9} \\
{\mybf $T_{\rm mid}^d$} & {\mybf 55074.99$\pm$0.03} & {\mybf 55087.721$_{-0.008}^{+0.007}$} & {\mybf 55081.57$\pm$0.02} & {\mybf 55083.88$\pm$0.03} & {\mybf 55079.07$\pm$0.01} \\
{\mybf $\Theta_{\rm e}^e$} & {\mybf {\mybf 15$^{+3}_{-2}$}} & {\mybf 31.7$^{+0.7}_{-0.8}$} & {\mybf 32$\pm$2} & {\mybf 34$^{+3}_{-2}$} & {\mybf 33.6$\pm$0.7} \\
\tableline
{\mybf $\chi^2_\nu$ (dof)} & {\mybf 1.18(80)} & {\mybf 0.97(80)} & {\mybf 1.08(80)} & {\mybf 1.01(80)} & {\mybf 1.03(80)} \\
\enddata
\tablecomments{\\*
$^a$ Units are 10$^{-3}$\,counts cm$^{-2}$ s$^{-1}$. \\*
$^b$ Refined orbital periods using an $O-C$ analysis.  Units are days. \\*
$^c$ The orbital period derivative at the 90$\%$ confidence interval found using an $O-C$ analysis.  Units are 10$^{-7}$\,d d$^{-1}$. \\*
$^d$ Units are BMJD. Phase 0 is defined as eclipse center. \\*
$^e$ Units are degrees.}
\label{Asymmetric Step and Ramp Model}
\end{deluxetable}

% Table 3.  This is the Eclipse Model Parameters
\begin{deluxetable}{ccccc}
\tablecolumns{5}
\tablewidth{0pc}
\tablecaption{{\mybf Eclipse Model Parameters, Including Historical Mid-Eclipse Times}}
\tablehead{
\colhead{{\mybf Model Parameter}} & \colhead{{\mybf IGR J18027-2016$^{a,c}$}} & \colhead{{\mybf IGR J18027-2016$^{a,d}$}} & \colhead{{\mybf XTE J1855-026$^{b,c}$}} & \colhead{{\mybf XTE J1855-026$^{b,d}$}}}
\startdata
{\mybf $\phi_{\rm ing}$} & {\mybf -0.147$_{-0.005}^{+0.004}$} & {\mybf -0.167$_{-0.004}^{+0.005}$} & {\mybf -0.131$_{-0.002}^{+0.001}$} & {\mybf -0.136$\pm$0.002} \\
{\mybf $\Delta{\phi}_{\rm ing}$} & {\mybf 0.053$_{-0.007}^{+0.005}$} & {\mybf 0.082$\pm$0.009} & {\mybf 0.038$\pm$0.002} & {\mybf 0.040$\pm$0.003} \\
{\mybf $\Delta{\phi}_{\rm eg}$} & {\mybf 0.053$_{-0.007}^{+0.005}$} & {\mybf 0.027$_{-0.006}^{+0.004}$} & {\mybf 0.038$\pm$0.002} & {\mybf 0.037$\pm$0.003} \\
{\mybf $C_{\rm ing}$$^e$} & {\mybf 1.59$\pm$0.05} & {\mybf 1.25$\pm$0.08} & {\mybf 2.64$\pm$0.06} & {\mybf 2.45$\pm$0.08} \\
{\mybf $C_{\rm eg}$$^e$} & {\mybf 1.59$\pm$0.05} & {\mybf 1.68$\pm$0.08} & {\mybf 2.64$\pm$0.06} & {\mybf 2.79$_{-0.08}^{+0.09}$} \\
{\mybf $\phi_{\rm egr}$} & {\mybf 0.099$\pm$0.003} & {\mybf 0.102$\pm$0.002} & {\mybf 0.094$\pm$0.002} & {\mybf 0.092$_{-0.002}^{+0.001}$} \\
{\mybf $C_{\rm ecl}$$^e$} & {\mybf 0.17$\pm$0.04} & {\mybf 0.14$\pm$0.04} & {\mybf -0.04$\pm$0.05} & {\mybf -0.03$\pm$0.05} \\
\tableline
{\mybf $\Delta{\phi}_{\rm ecl}$} & {\mybf 0.193$^{+0.007}_{-0.009}$} & {\mybf 0.19$\pm$0.01} & {\mybf 0.187$\pm$0.003} & {\mybf 0.187$\pm$0.004} \\
{\mybf $P_{\rm orb}^f$} & {\mybf 4.56982$\pm$0.00003} & {\mybf 4.56993$\pm$0.00003} & {\mybf 6.07412$\pm$0.00003} & {\mybf 6.07414$\pm$0.00003} \\
{\mybf $\dot{P}_{\rm orb}^g$} & {\mybf 0.8$\pm$0.9} & {\mybf 0.2$\pm$1.1} & {\mybf -0.1$\pm$0.5} & {\mybf 0.0$\pm$0.5} \\
{\mybf $T_{\rm mid}^h$} & {\mybf 55083.78$\pm$0.01} & {\mybf 55083.82$\pm$0.01} & {\mybf 55079.056$\pm$0.009} & {\mybf 55079.07$\pm$0.01} \\
{\mybf $\Theta_{\rm e}^i$} & {\mybf 35$\pm$1} & {\mybf 34$\pm$2} & {\mybf 33.6$^{+0.5}_{-0.6}$} & {\mybf 33.7$\pm$0.7} \\
\tableline
{\mybf $\chi^2_\nu$ (dof)} & {\mybf 1.20(77)} & {\mybf 0.93(80)} & {\mybf 1.02(77)} & {\mybf 1.05(80)} \\
\enddata
\tablecomments{\\*
$^a$ Includes the mid-eclipse times derived in \citet{2005A&A...439..255H}, \citet{2009RAA.....9.1303J} and \citet{2015A&A...577A.130F}. \\*
$^b$ Includes the mid-eclipse times derived in \citet{2002ApJ...577..923C} and \citet{2015A&A...577A.130F}. \\*
$^c$ Assuming a Symmetric Eclipse Profile. \\*
$^d$ Assuming an Asymmetric Eclipse Profile. \\*
$^e$ Units are 10$^{-3}$\,counts cm$^{-2}$ s$^{-1}$. \\*
$^f$ Refined orbital periods using an $O-C$ analysis.  Units are days. \\*
$^g$ The orbital period derivative at the 90$\%$ confidence interval found using an $O-C$ analysis.  Units are 10$^{-7}$\,d d$^{-1}$. \\*
$^h$ Units are BMJD. Phase 0 is defined as eclipse center. \\*
$^i$ Units are degrees.}
\label{Historic Step and Ramp Model}
\end{deluxetable}

\begin{figure}[ht]
\centerline{\includegraphics[width=3in]{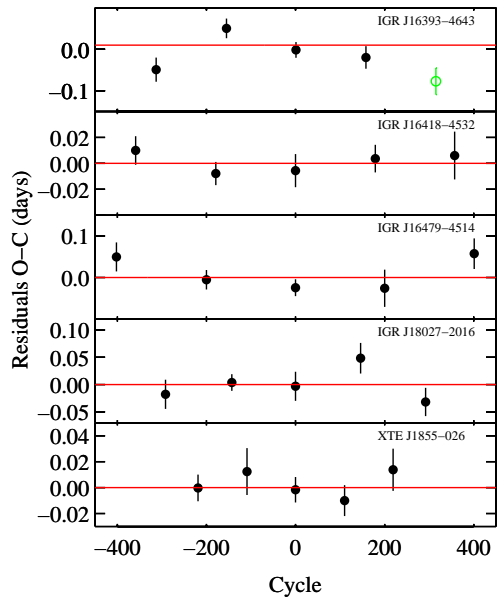}}
\figcaption[april14-2015_BAT-OCcorrections_allsources.ps]{
The observed minus calculated ($O-C$) eclipse time residuals for IGR J16393-4643 (top), IGR J16418-4532 (second panel), IGR J16479-4514 (middle), IGR J18027-2016 (fourth panel) and XTE J1855-026 (bottom) {\mybf fit using a symmetric step-and-ramp function}.  We subtract the best linear polynomial fit for each source and correct the orbital periods accordingly.  For IGR J16393-4643 (top) we only use the first four points to obtain a good fit (see Section~\ref{IGR J16393-4643 Results}).
\label{O-C Residuals}
}
\end{figure}

\begin{figure}[ht]
\centerline{\includegraphics[width=3in]{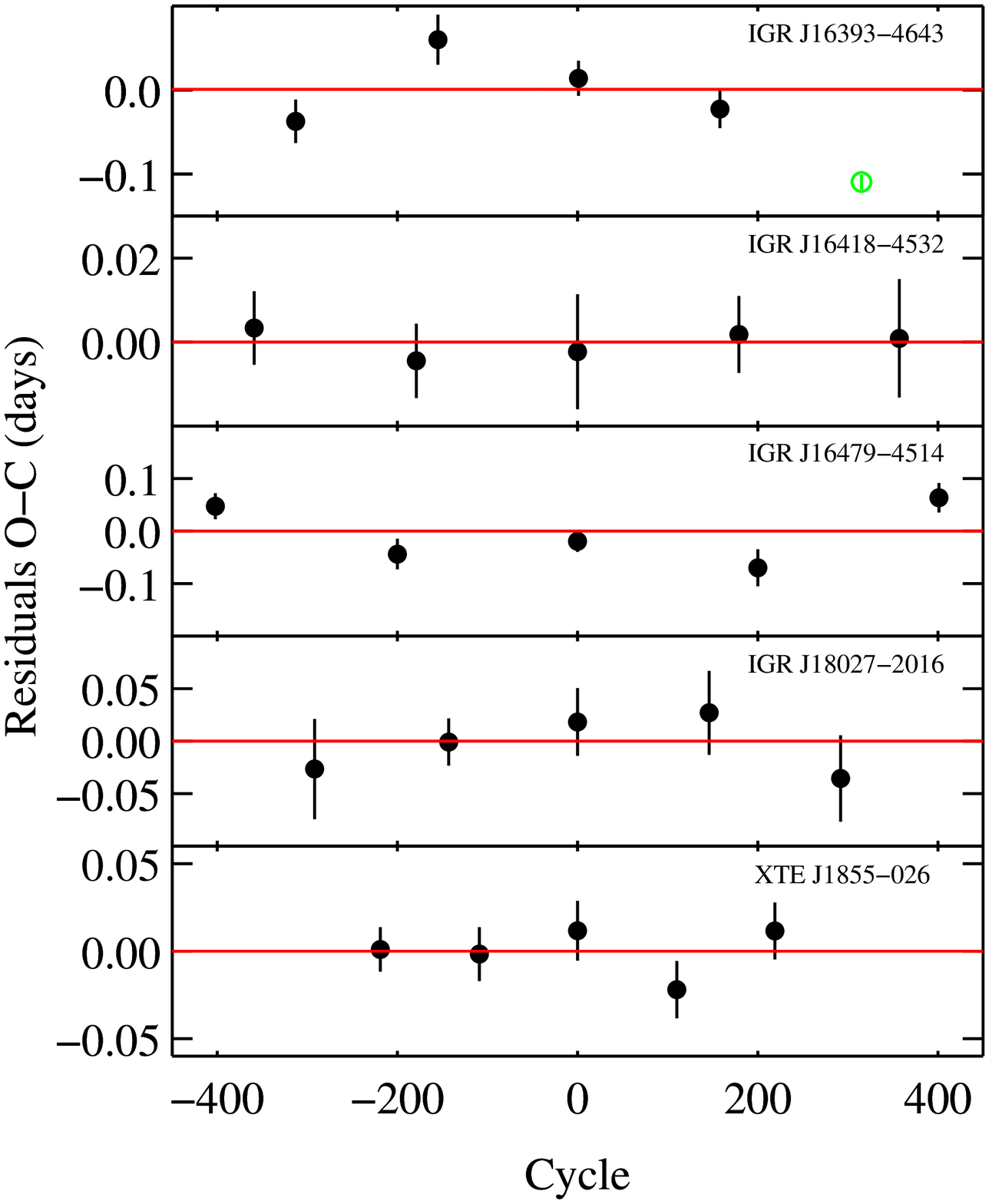}}
\figcaption[april14-2015_BAT-OCasymmetric_allsources.ps]{
The observed minus calculated ($O-C$) eclipse time residuals for IGR J16393-4643 (top), IGR J16418-4532 (second panel), IGR J16479-4514 (middle), IGR J18027-2016 (fourth panel) and XTE J1855-026 (bottom) fit {\mybf using an asymmetric step-and-ramp function}.  We subtract the best linear polynomial fit for each source and correct the orbital periods accordingly.  For IGR J16393-4643 (top) we only use the first four points to obtain a good fit (see Section~\ref{IGR J16393-4643 Results}).
\label{O-C Asymmetric}
}
\end{figure}

\begin{figure}[ht]
\centerline{\includegraphics[width=3in]{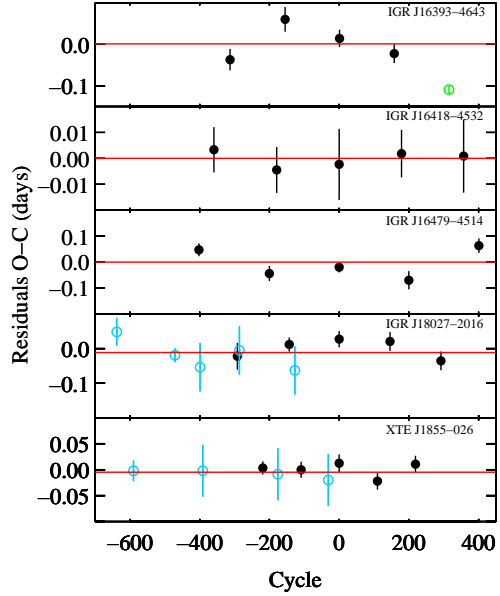}}
\figcaption[may4-2015_BAT-OCcombined_allsources.ps]{
Derived $O-C$ eclipse time residuals for IGR J16393-4643 (top), IGR J16418-4532 (second panel), IGR J16479-4514 (middle), IGR J18027-2016 (fourth panel) and XTE J1855-026 (bottom) fit combined with those in the literature.  We subtract the best linear polynomial fit for each source and correct the orbital periods accordingly.  For IGR J16393-4643 (top) we only use the first four points to obtain a good fit (see Section~\ref{IGR J16393-4643 Results}).
\label{O-C Historic}
}
\end{figure}

The eclipse duration, time of mid-eclipse, and eclipse half-angle ($\Theta_{\rm e}=\Delta{\phi}_{\rm ecl}$$\times$180$\degr$) from fitting the BAT folded light curves for each source are reported in Tables~\ref{Step and Ramp Model}--~\ref{Historic Step and Ramp Model}.  For each source, we initially used an ephemeris based on our determination of the orbital period from the DFT and time of mid-eclipse.  Using an `observed minus calculated' $O-C$ analysis (see {\mybf Figures~\ref{O-C Residuals}--\ref{O-C Historic}}), we refined the orbital periods and improved on the time of mid-eclipse for each XRB in our sample.  {\mybf We note that no eclipses are visible in the unfolded light curves and it is necessary to observe multiple cycles of folded light curves in order for eclipses to be seen. We} divide the light curves {\mybf into} five equal time intervals ($\sim$670\,days), with the exception of IGR J16393-4643 (see Section~\ref{IGR J16393-4643 Results}), and calculate the mid-eclipse epoch for each interval {\mybf (see Table~\ref{O-C Table}) In the cases of IGR J18027-2016 and XTE J1855-026, we combine our derived mid-eclipse times with those reported in the literature \citep[][and references therein]{2015A&A...577A.130F}.  We note that while mid-eclipse times were previously derived for both IGR J16393-4643 and IGR J16479-4514 (see Table~\ref{Historic O-C Table}), these were not used since no error estimate was reported \citep{2015MNRAS.446.4148I,2009AIPC.1126..319B}.}  We then fit the mid-eclipse times using the orbital change function (see Equation~\ref{Orbital Change Function}) where $n$ is the number of binary orbits {\mybf given to the nearest integer}, $P_{\rm orb}$ is the orbital period in days, $\dot{P}_{\rm orb}$ is the period derivative at $T_0$, and the error on the linear term is the orbital period error.  In all five cases, we improve the error estimate on the orbital period by nearly an order of magnitude (see Section~\ref{Five Eclipsing HMXBs}).  We do not find a significant $\dot{P}_{\rm orb}$ for any source {\mybf (see Tables~\ref{Step and Ramp Model}--~\ref{Historic Step and Ramp Model})}.

\begin{equation}
\label{Orbital Change Function}
T_{\rm n}=T_0+n P_{\rm orb}+\frac{1}{2} n^2 P_{\rm orb} \dot{P}_{\rm orb}
\end{equation}

%figure 1

%figure 2

%figure 3

% Table 4
\begin{deluxetable}{ccccc}
\tablecolumns{5}
\tabletypesize{\small}
\tablewidth{0pc}
\tablecaption{{\mybf Mid-eclipse Time Measurements For $O-C$ Analysis}}
\tablehead{
\colhead{{\mybf Source}} & \colhead{{\mybf Orbital}} & \colhead{{\mybf Mid-eclipse time$^a$}} & \colhead{{\mybf Mid-eclipse time$^b$}} \\
\colhead{} & \colhead{{\mybf Cycle (N)}} & \colhead{{\mybf (MJD)}} & \colhead{{\mybf (MJD)}}}
\startdata
{\mybf IGR J16393-4643} & {\mybf -313} & {\mybf 53748.46$\pm$0.03} & {\mybf 53748.44$\pm$0.03} \\
{\mybf IGR J16393-4643} & {\mybf -155} & {\mybf 54418.15$_{-0.02}^{+0.04}$} & {\mybf 54418.14$_{-0.03}^{+0.04}$} \\
{\mybf IGR J16393-4643} & {\mybf 1} & {\mybf 55079.22$_{-0.02}^{+0.07}$} & {\mybf 55079.22$_{-0.02}^{+0.03}$} \\
{\mybf IGR J16393-4643} & {\mybf 158} & {\mybf 55744.55$_{-0.03}^{+0.02}$} & {\mybf 55744.54$_{-0.08}^{+0.02}$} \\
{\mybf IGR J16393-4643} & {\mybf 315} & {\mybf \nodata$^c$} & {\mybf \nodata$^c$} \\
\tableline
{\mybf IGR J16418-4532} & {\mybf -359} & {\mybf 53745.50$\pm$0.01} & {\mybf 53745.491$_{-0.009}^{+0.013}$} \\
{\mybf IGR J16418-4532} & {\mybf -179} & {\mybf 54418.461$^{+0.009}_{-0.010}$} & {\mybf 54418.468$\pm$0.009} \\
{\mybf IGR J16418-4532} & {\mybf 0} & {\mybf 55087.71$\pm$0.01} & {\mybf 55087.72$\pm$0.01} \\
{\mybf IGR J16418-4532} & {\mybf 179} & {\mybf 55756.963$^{+0.011}_{-0.008}$} & {\mybf 55756.967$\pm$0.009} \\
{\mybf IGR J16418-4532} & {\mybf 357} & {\mybf 56422.47$\pm$-0.02} & {\mybf 56422.47$_{-0.01}^{+0.02}$} \\
\tableline
{\mybf IGR J16479-4514} & {\mybf -402} & {\mybf 53747.19$_{-0.03}^{+0.02}$} & {\mybf 53747.21$_{-0.03}^{+0.02}$} \\
{\mybf IGR J16479-4514} & {\mybf -200} & {\mybf 54417.63$\pm$0.02} & {\mybf 54417.61$_{-0.02}^{+0.03}$} \\
{\mybf IGR J16479-4514} & {\mybf 0} & {\mybf 55081.47$^{+0.03}_{-0.02}$} & {\mybf 55081.48$_{-0.03}^{+0.02}$} \\
{\mybf IGR J16479-4514} & {\mybf 200} & {\mybf 55745.33$_{-0.04}^{+0.03}$} & {\mybf 55745.28$\pm$0.04} \\
{\mybf IGR J16479-4514} & {\mybf 401} & {\mybf 56412.60$_{-0.04}^{+0.06}$} & {\mybf 56412.59$_{-0.04}^{+0.03}$} \\
\tableline
{\mybf IGR J18027-2016} & {\mybf -292} & {\mybf 53749.41$_{-0.03}^{+0.05}$} & {\mybf 53749.40$_{-0.05}^{+0.04}$} \\
{\mybf IGR J18027-2016} & {\mybf -143} & {\mybf 54430.32$\pm$0.02} & {\mybf 54430.34$_{-0.02}^{+0.04}$} \\
{\mybf IGR J18027-2016} & {\mybf 0} & {\mybf 55083.79$_{-0.03}^{+0.02}$} & {\mybf 55083.85$_{-0.03}^{+0.08}$} \\
{\mybf IGR J18027-2016} & {\mybf 146} & {\mybf 55751.02$\pm$0.03} & {\mybf 55751.06$_{-0.04}^{+0.03}$} \\
{\mybf IGR J18027-2016} & {\mybf 292} & {\mybf 56418.12$_{-0.03}^{+0.02}$} & {\mybf 56418.20$_{-0.04}^{+0.03}$} \\
\tableline
{\mybf XTE J1855-026} & {\mybf -219} & {\mybf 53748.82$\pm$0.01} & {\mybf 53748.84$\pm$0.01} \\
{\mybf XTE J1855-026} & {\mybf -109} & {\mybf 54416.99$\pm$0.02} & {\mybf 54416.99$\pm$0.02} \\
{\mybf XTE J1855-026} & {\mybf 0} & {\mybf 55079.05$\pm$0.01} & {\mybf 55079.09$_{-0.03}^{+0.02}$} & \\
{\mybf XTE J1855-026} & {\mybf 110} & {\mybf 55747.20$\pm$0.01} & {\mybf 55747.21$\pm$0.02} \\
{\mybf XTE J1855-026} & {\mybf 219} & {\mybf 56409.30$\pm$0.02} & {\mybf 56409.33$\pm$0.02} \\
\enddata
\tablecomments{{\mybf $^a$ Obtained using the symmetric step-and-ramp function. \\*
$^b$ Obtained using the asymmetric step-and-ramp function. \\*
$^c$ A bad fit for IGR J16393-4643 was obtained in the mid-eclipse time between MJD\,56079--56745.}}
\label{O-C Table}
\end{deluxetable}

% Table 5
\begin{deluxetable}{cccccc}
\tablecolumns{6}
\tabletypesize{\small}
\tablewidth{0pc}
\tablecaption{{\mybf Historic Mid-eclipse Time Measurements For IGR J18027-2016 and XTE J1855-026}}
\tablehead{
\colhead{{\mybf Source}} & \colhead{{\mybf Orbital}} & \colhead{{\mybf Mid-eclipse time}} & \colhead{{\mybf Satellite}} & \colhead{{\mybf Reference}} \\
\colhead{} & \colhead{{\mybf Cycle (N)}} & \colhead{{\mybf (MJD)}} & \colhead{{\mybf (MJD)}} & \colhead{} & \colhead{}}
\startdata
{\mybf IGR J16393-4643} & {\mybf -391} & {\mybf 53417.955} & {\mybf \textsl{Swift}} & {\mybf \citet{2015MNRAS.446.4148I}} \\
{\mybf IGR J16479-4514} & {\mybf -161} & {\mybf 54547.05418} & {\mybf \textsl{Swift}} & {\mybf \citet{2009AIPC.1126..319B}} \\
\tableline
{\mybf IGR J18027-2016} & {\mybf -638} & {\mybf 52168.22$\pm$0.12} & {\mybf \textsl{BeppoSAX}} & {\mybf \citet{2003ApJ...596L..63A}} \\
{\mybf IGR J18027-2016} & {\mybf -638} & {\mybf 52168.26$\pm$0.04} & {\mybf \textsl{BeppoSAX}} & {\mybf \citet{2005A&A...439..255H}} \\
{\mybf IGR J18027-2016} & {\mybf -471} & {\mybf 52931.37$\pm$0.04} & {\mybf \textsl{INTEGRAL}} & {\mybf \citet{2005A&A...439..255H}} \\
{\mybf IGR J18027-2016} & {\mybf -399} & {\mybf 53260.37$\pm$0.07} & {\mybf \textsl{INTEGRAL}} & {\mybf \citet{2009RAA.....9.1303J}} \\
{\mybf IGR J18027-2016} & {\mybf -286} & {\mybf 53776.82$\pm$0.07} & {\mybf \textsl{Swift}} & {\mybf \citet{2009RAA.....9.1303J}} \\
{\mybf IGR J18027-2016} & {\mybf -267} & {\mybf 53863.10$\pm$0.14} & {\mybf \textsl{INTEGRAL}} & {\mybf \citet{2015A&A...577A.130F}} \\
{\mybf IGR J18027-2016} & {\mybf -127} & {\mybf 54503.38$\pm$0.07} & {\mybf \textsl{Swift}} & {\mybf \citet{2009RAA.....9.1303J}} \\
\tableline
{\mybf XTE J1855-026} & {\mybf -590} & {\mybf 51495.25$\pm$0.02} & {\mybf \textsl{RXTE}} & {\mybf \citet{2002ApJ...577..923C}} \\
{\mybf XTE J1855-026} & {\mybf -391} & {\mybf 52704.04$\pm$0.05} & {\mybf \textsl{INTEGRAL}} & {\mybf \citet{2015A&A...577A.130F}} \\
{\mybf XTE J1855-026} & {\mybf -176} & {\mybf 54009.97$\pm$0.05} & {\mybf \textsl{INTEGRAL}} & {\mybf \citet{2015A&A...577A.130F}} \\
{\mybf XTE J1855-026} & {\mybf -31} & {\mybf 54890.68$\pm$0.05} & {\mybf \textsl{INTEGRAL}} & {\mybf \citet{2015A&A...577A.130F}} \\
\enddata
\tablecomments{{\mybf Historical mid-eclipse times for IGR J16479-4514, IGR J18027-2016 and XTE J1855-026 found using \textsl{RXTE}, \textsl{Swift} BAT and \textsl{INTEGRAL}.}}
\label{Historic O-C Table}
\end{deluxetable}

X-ray binaries that are eclipsing have an eclipse duration that is only dependent on the radius of the mass donor, inclination angle of the system and the orbital separation of the components provided that the orbit is circular ($e=0$).  Using the observed orbital period and Kepler's third law, the duration can be written in terms of the sum of the donor star and compact object masses, which stipulates that the eclipse half-angle, $\Theta_{\rm e}$, can now be expressed in terms of the radius, inclination and masses of the components.  In one set of calculations, we assume a 1.4\,$M_\sun$ compact object which may be appropriate for an accreting neutron star \citep{1931ApJ....74...81C}.  The region allowed by the measured eclipse half-angle for each binary in Mass-Radius space is shown in Section~\ref{Five Eclipsing HMXBs}.  The inclination is constrained between edge-on orbits (left boundary of the dark shaded region) and close to face-on orbits (the right boundary of the light shaded region).  We can attach additional constraints assuming that the mass donor underfills the Roche-lobe (right boundary of the dark shaded region), which is dependent on the mass ratio of the system and the orbital separation.  To calculate the eclipse half-angle and the Roche-lobe radius, we used Equation 7 in {\mybf \citet{1984ARA&A..22..537J}, also used by \citet{1996ApJ...459..259R},} and Equation 2 in \citet{1983ApJ...268..368E}, respectively.  Further constraints on the parameters of the donor star are imposed with pulse-timing techniques (dashed red lines in Figures~\ref{J18027 Mass Radius Plot} and~\ref{J1855 Mass Radius Plot}).  For the systems where pulse-timing results were not available, we additionally calculated the minimum inclination angle of the system, $i_{\rm min}$, that is consistent with the measured eclipse half-angle (see Table~\ref{Step and Ramp Model}).

When the {\mybf semi-amplitude of the} radial velocities of both the compact object and the mass donor are known (e.g. IGR J18027-2016, {\mybf XTE J1855-026}), the mass ratio between the compact object and mass donor can be calculated {\mybf \citep[Equation 6,][]{1984ARA&A..22..537J}}.  This means that in addition to the radius and mass of the donor star, the mass of the compact object can be constrained.  The mass of the donor star can be written in terms of the {\mybf semi-amplitude of the} radial velocity of the compact object, orbital period, Newton's gravitational constant, inclination angle of the system and the mass ratio {\mybf \citep{1984ARA&A..22..537J}}.  Likewise, the compact object mass can be written in terms of the {\mybf semi-amplitude of the} of the radial velocity of the donor star, orbital period, Newton's gravitational constant, inclination angle of the system and the mass ratio {\mybf \citep{1984ARA&A..22..537J}}.  To calculate the masses of both the donor star and the compact object, we used Equations 2 and 3 in {\mybf \citet{1999MNRAS.307..357A}, also used by \citet{1983adsx.conf...13R}}.

For consistency, we compare our derived constraints on the masses and radii of the donor stars with those expected for the previously proposed spectral types.  For the systems where pulse-timing results were not available, we calculate the predicted eclipse half-angles as a function of inclination angle using the mass and radius for the derived spectral types (see Section~\ref{Five Eclipsing HMXBs}).  Generally we used results from \citet{2006ima..book.....C} for main-sequence, giant and supergiant luminosity classes.  These are represented by the red, green and blue dashed lines in mass-radius space for each system (see Section~\ref{Five Eclipsing HMXBs}). For O-type supergiants, we also use Tables 3 and 6 in \citet{2005A&A...436.1049M} to compare our results (see blue dotted lines in Figures~\ref{J16393 Mass Radius Plot},~\ref{J16418 Mass Radius Plot} and~\ref{J16479 Mass Radius Plot}).  The constraints for B-type supergiants are additionally compared with Tables 3 and 6 in \citet{2007A&A...463.1093L} for B-type supergiants (blue crosses in Figures~\ref{J18027 Mass Radius Plot} and~\ref{J1855 Mass Radius Plot}).

\section{Five Eclipsing HMXBs}
\label{Five Eclipsing HMXBs}

\subsection{IGR J16393-4643 (=AX J16390.4-4642)}
\label{IGR J16393-4643 Results}

IGR J16393-4643 is an HMXB first discovered and listed as AX J16390.4-4642 in the \textsl{ASCA} Faint Source Catalog \citep{2001ApJS..134...77S} and was later detected with \textsl{INTEGRAL} \citep{2006A&A...447.1027B}.  The average flux in the 20--40\,keV band was found to be 5.1$\times$10$^{-11}$\,erg cm$^{-2}$ s$^{-1}$, and intensity variations were found to exceed a factor of 20 \citep{2006A&A...447.1027B}.  In the 2--10\,keV energy band, the unabsorbed flux was found to be 9.2$\times$10$^{-11}$\,erg cm$^{-2}$ s$^{-1}$ \citep{2006A&A...447.1027B}.  A proposed mass donor 2MASS J16390535-4242137 was found in the \textsl{XMM-Newton} error circle, which is thought to be an OB supergiant \citep{2006A&A...447.1027B}.  However, a precise position of the donor star obtained with \textsl{Chandra} shows this candidate to be positionally inconsistent with the X-ray source \citep{2012ApJ...751..113B}. Using the \textsl{Spitzer} Galactic Legacy Infrared Mid-Plane Survey (GLIMPSE), \citet{2012ApJ...751..113B} proposed that the counterpart must be a distant reddened B-type main-sequence star.

Using \textsl{INTEGRAL} and \textsl{XMM-Newton}, \citet{2006A&A...447.1027B} found a 912.0$\pm$0.1\,s modulation, which was interpreted as the neutron star rotation period.  \citet{2015MNRAS.446.4148I} recently refined this to 908.79$\pm$0.01\,s using \textsl{Suzaku}.  A $\sim$3.7\,day orbital period was suggested using a pulse timing analysis, although orbital periods of $\sim$50.2 and $\sim$8.1\,days were not completely ruled out \citep{2006ApJ...649..373T}.  While various possible orbital solutions and accretion mechanisms have been proposed, orbital periods of 4.2368$\pm$0.0007 and 4.2371$\pm$0.0007\,days were clearly found from data from \textsl{Swift}-BAT and \textsl{RXTE} PCA, respectively \citep{2010ATel.2570....1C}.  This was refined to 4.2386$\pm$0.0003\,d \citep{2013ApJ...778...45C}, also using BAT data. \citet{2015MNRAS.446.4148I} recently derived an orbital period of $\sim$366150\,s (4.24\,d) with BAT, which is also consistent with the result from \citet{2013ApJ...778...45C}. The position in Corbet's diagram shows that IGR J16393-4643 is an SGXB \citep{2013ApJ...778...45C}.  \citet{2013ApJ...778...45C} identified the presence of a possible superorbital period of $\sim$15\,days; although with low significance.

%figure 4
\begin{figure}[ht]
\centerline{\includegraphics[angle=-90,width=3in]{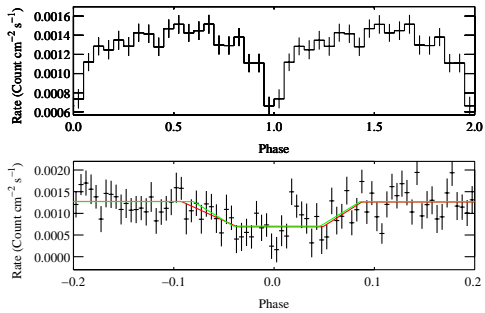}}
\figcaption[4-22-15_BAT_proper-folded-lightcurves_IGRJ16393-4643.ps]{
\textsl{Swift}-BAT light curve of IGR J16393-4643 {\mybf in the 15--50\,keV band} folded on the orbital period (top) using 20 bins.  T0 is defined at BMJD\,55074.99, corresponding to mid-eclipse.  A detailed folded light curve with 80 bins (bottom) is fit with {\mybf both a symmetric ``step and ramp" function (green) and asymmetric ``step and ramp" function (red)}, which model the eclipse.  The symmetric ``step and ramp" function was shifted accordingly.
\label{J16393 Folded Half Angle}
}
\end{figure}

The BAT light curves folded on the orbital period revealed a sharp dip, which was interpreted as an eclipse \citep{2013ApJ...778...45C,2015MNRAS.446.4148I}.  With \textsl{Swift} BAT, \citet{2015MNRAS.446.4148I} constrained the eclipse half angle to be $\sim$17$\degr$, corresponding to a duration of $\sim$0.75\,d ($\sim$65.1\,ks).  Using the relationship between eclipse duration and stellar radius, along with the definition of the Roche-lobe from \citet{1984avis.book.....B}, \citet{2015MNRAS.446.4148I} calculated the allowed range of orbital inclinations of the system.  Assuming a star with spectral type O9 I \citep{2012ApJ...751..113B}, the orbital inclination was constrained to 39--57$\degr$ \citep{2015MNRAS.446.4148I}.  A main-sequence B-type star yields orbital inclinations between 60--77$\degr$ \citep{2012ApJ...751..113B,2015MNRAS.446.4148I}.

We derive a {\mybf 4.2378$\pm$0.0004\,d} orbital period for IGR J16393-4643 using a DFT, which is consistent with the results from \citet{2013ApJ...778...45C}.  Using an $O-C$ analysis (see Section~\ref{Eclipse Modeling}), this is further refined to {\mybf 4.23810$\pm$0.00007\,d}.  We note that we obtain a bad fit in the mid-eclipse time between MJD\,56079--56745.  As a result, we only use data between MJD\,53416--56078 in our $O-C$ analysis (see {\mybf Figures~\ref{O-C Residuals}--\ref{O-C Historic}}--~\ref{O-C Historic}). {\mybf Using the quadratic orbital change function (see Equation~\ref{Orbital Change Function}), we find the orbital period derivative to be -5$\pm$4$\times$10$^{-7}$\,d d$^{-1}$, which is consistent with zero.}  The duration of the observed eclipse was calculated to be {\mybf 31$^{+6}_{-5}$\,ks} {\mybf (0.36$^{+0.07}_{-0.06}$\,d)}, yielding an eclipse half-angle of 15$_{-3}^{+2}$$\degr$ (see Table~\ref{Step and Ramp Model}).  We find these to be consistent with the result from \citet{2015MNRAS.446.4148I}.  The source flux does not reach 0\,counts cm$^{-2}$ s$^{−1}$ in the folded light curves during eclipse (see Figure~\ref{J16393 Folded Half Angle}).  We interpret this dip as an eclipse since the feature is persistent over many years of data.  The rapid ingress and egress requires obscuration by clearly defined boundaries that are suggestive of an object such as the mass donor in the system {\mybf \citep[e.g.][]{2014ApJ...793...77C}}.  We discuss the nature of the non-zero flux during eclipse in Section~\ref{What is the nature of the non-zero eclipse flux in IGR J16393-4643?}.

\begin{figure}[ht]
\centerline{\includegraphics[width=3in]{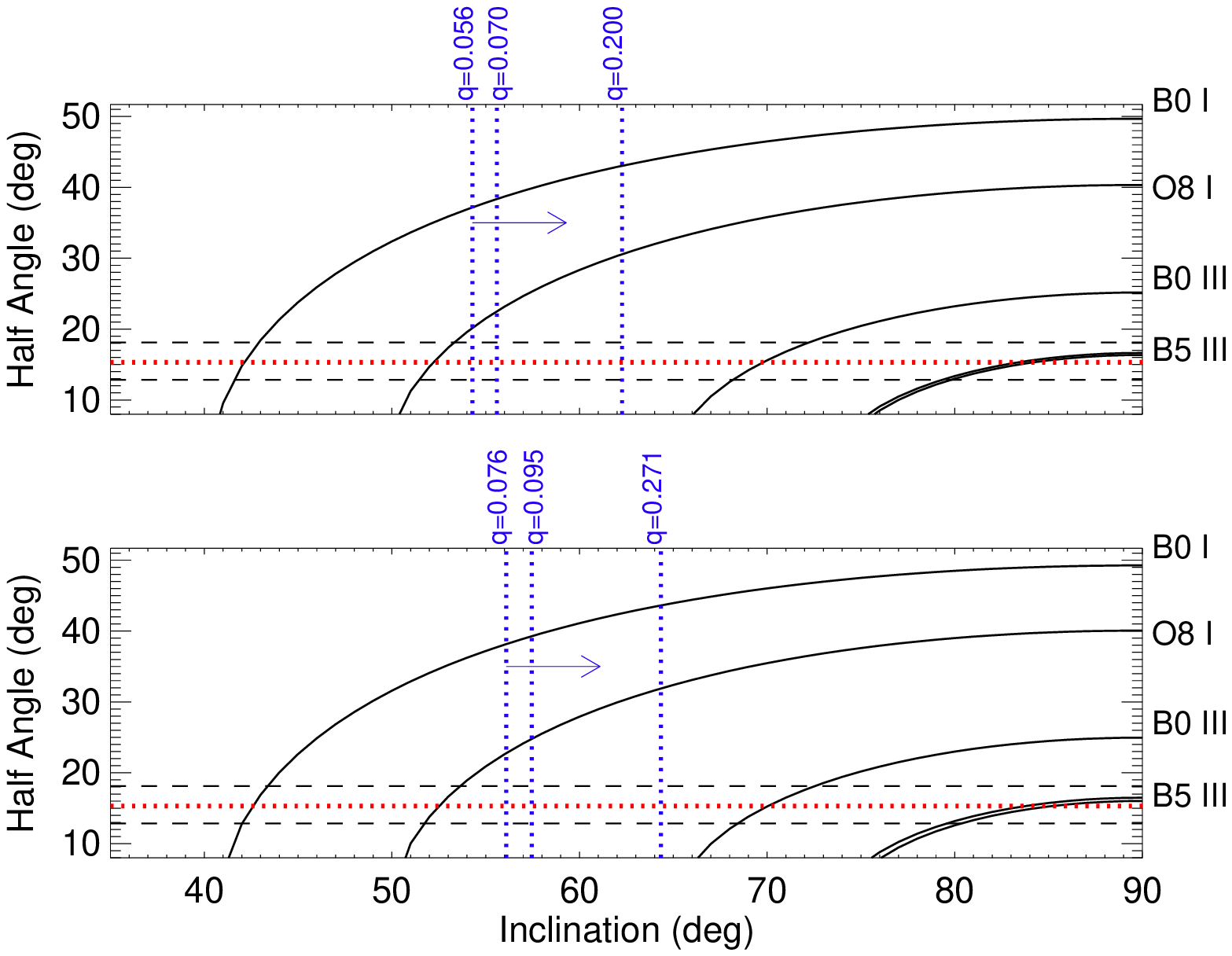}}
\figcaption[4-21-2015_IGRJ16393-4643_weighted_eclipse_param.eps]{
The black curves show the predicted eclipse half angle of IGR J16393-4643 as a function of inclination angle for stars with the indicated spectral types.  The red and black dashed lines indicate the eclipse half angle and estimated error as measured by \textsl{Swift} BAT.  We assume a neutron star mass of 1.4\,$M_\sun$ (top) and of mass 1.9\,$M_\sun$ (bottom) and typical masses and radii for the assumed companion spectral type (see Table~\ref{J16393 Primary Parameters}).  The blue vertical dashed lines indicate the lower limit of the inclination angle.  Inclinations to the left of these correspond to stars that overfill the Roche-lobe.
\label{J16393 Eclipse Half Angle}
}
\end{figure}

We calculate the predicted eclipse half-angle $\Theta_{\rm e}$ as a function of inclination angle of the system (see Figure~\ref{J16393 Eclipse Half Angle}).  The calculation assumes a neutron star mass of 1.4\,$M_\sun$, and the primary stellar masses and radii given in Table~\ref{J16393 Primary Parameters}.  We calculate the minimum inclination angle of the system, $i_{\rm min}$, that is consistent with the measured eclipse half-angle (see Table~\ref{J16393 Primary Parameters}).  We find that stars with spectral types B0 V, B0-5 III and B0 I satisfy the constraint imposed by the eclipse half-angle (see Table~\ref{J16393 Primary Parameters}).  We note that while a B5 III star satisfies the constraint imposed by the minimum value of the eclipse half-angle under the assumption that the neutron star is 1.4\,$M_\sun$, this spectral type does not satisfy the eclipse half-angle for a more massive neutron star.

%figure 5

% Table 6
\begin{deluxetable}{ccccccccc}
\tablecolumns{9}
\tablewidth{0pc}
\tablecaption{Physical Parameters for Previously Proposed Mass Donors for IGR J16393-4643}
\tablehead{
\colhead{Spectral Type} & \colhead{$M/M_\sun$} & \colhead{$q$$^a$} & \colhead{$R/R_\sun$} & \colhead{$R_{\rm L}$$/R_\sun$$^b$} & \colhead{$i_{\rm min}$$\degr$$^c$}} 
\startdata
\textsl{B0 III} & \textsl{20} & 0.070 & \textsl{13} & 18.5 & {\mybf 68} \\
\textsl{B5 III} & \textsl{7} & 0.200 & \textsl{6.3} & 11.7 & {\mybf 79} \\
\textsl{B0 V} & \textsl{17.5} & 0.080 & \textsl{8.4} & 17.5 & {\mybf 79} \\
\textsl{B0 I} & \textsl{25} & 0.056 & \textsl{25$^d$} & 20.4$^d$ & {\mybf 41} \\
\enddata
\tablecomments{The values in italics are obtained from \citet{2006ima..book.....C} \\*
$^a$ The mass ratio, $q$, is defined as $M_{\rm x}/M_{\rm c}$ where $M_{\rm x}$ is the compact object and $M_{\rm c}$ is the donor star. \\*
$^b$ The definition for the Roche lobe, $R_{\rm L}$, as given in \citet{1983ApJ...268..368E}, assuming $M_{\rm NS}$ is 1.4\,$M_\sun$. \\*
$^c$ The minimum inclination angle of the system that is consistent with the measured eclipse half-angle. \\*
$^d$ A B0 I classification significantly overfills the Roche-lobe and are therefore, it is excluded from our analysis. \\*
}
\label{J16393 Primary Parameters}
\end{deluxetable}

%figure 6
\begin{figure}[ht]
\centerline{\includegraphics[width=3in]{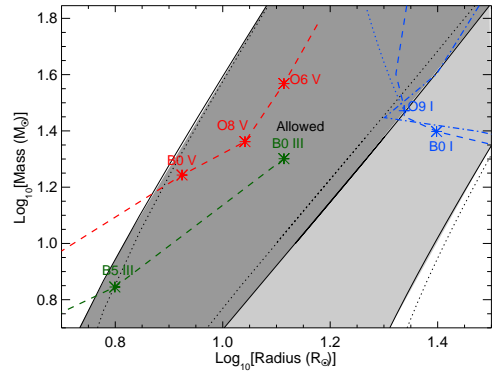}}
\figcaption[april22-2015_IGRJ16393-4643_mass-constraints.eps]{
Log-log plot of stellar mass as a function of stellar radius for IGR J16393-4643.  The dark shaded region indicates the allowed spectral types that satisfy the constraints imposed by both the eclipse duration and Roche-lobe size.  The light shaded region only indicates spectral types that satisfy the observed eclipse duration.  Stellar masses and radii are reported in Table~\ref{J16393 Primary Parameters}.  The red, green and blue lines indicate interpolations for main-sequence, giant and supergiant luminosity classes, respectively.  The dashed, dotted and dash-dotted lines indicate stellar masses and radii derived from \citet{2006ima..book.....C}, \citet{2005A&A...436.1049M} and \citet{2000asqu.book.....A}, respectively.
\label{J16393 Mass Radius Plot}
}
\end{figure}

Since the X-ray luminosity is lower than what would be expected for Roche-lobe overflow \citep{2004RMxAC..21..128K}, we can attach an additional constraint assuming that the mass donor underfills the Roche-lobe radius (see Figure~\ref{J16393 Mass Radius Plot}).  We note that while the derived masses and radii for the spectral type B0 I from \citet{2006ima..book.....C} satisfies the eclipse half-angle, the assumed radius would be larger than the Roche-lobe radius \citep{1983ApJ...268..368E}.  Therefore, an B0 I spectral type must be excluded (see Table~\ref{J16393 Primary Parameters}).

\subsection{IGR J16418-4532}

IGR J16418-4532 is a candidate Supergiant Fast X-ray Transient (SFXT) first discovered with the \textsl{INTEGRAL} satellite by \citet{2004ATel..224....1T} at a flux of 3$\times$10$^{-11}$\,erg cm$^{-2}$ s$^{-1}$ in the 20--40\,keV band.  The near-infrared spectral energy distribution of the most probable Two Micron All Sky Survey (2MASS) counterpart was measured with the 3.5\,m New Technology Telescope (NTT) at La Silla Observatory, \citet{2008A&A...484..801R} found a spectral type of O8.5 I.  \citet{2013A&A...560A.108C} proposed a spectral type of BN0.5 Ia based on features in the near-infrared spectrum such as Br$(7-4)$ and the emission and absorption of neutral helium.  Using \textsl{XMM-Newton}, \citet{2006A&A...453..133W} found a 1246$\pm$100\,s modulation, which was interpreted as a neutron star rotation period.  This was later refined to 1212$\pm$6\,s \citep{2012MNRAS.420..554S}, also using \textsl{XMM-Newton}.  Recently, \citet{2013MNRAS.433..528D} further refined the rotation period to 1209.1$\pm$0.4\,s with \textsl{XMM-Newton}.  A $\sim$3.73\,d orbital period was found using data from the \textsl{Swift} BAT and the \textsl{RXTE} ASM instruments, where $P_{\rm orb}$ was reported as 3.753$\pm$0.004\,d and 3.7389$\pm$0.0004\,d, respectively \citep{2006ATel..779....1C}.  Using an extended ASM dataset, \citet{2011ApJS..196....6L} found a 3.73886$\pm$0.00028\,d period, which is consistent with the earlier result from \citet{2006ATel..779....1C}.  \citet{2013ApJ...778...45C} further refined this to 3.73886$\pm$0.00014\,d using BAT.  A $\sim$14.7\,d modulation was found using BAT and \textsl{INTEGRAL}, which was interpreted as a superorbital period \citep{2013ApJ...778...45C, 2013ATel.5131....1D}.

The ASM and BAT light curves folded on the orbital period revealed a sharp dip with near zero mean flux, which was interpreted as a total eclipse \citep{2006ATel..779....1C}.  Subsequent observations of the eclipse included \textsl{Swift} X-ray Telescope (XRT) \citep{2012MNRAS.419.2695R} and \textsl{INTEGRAL} \citep{2013MNRAS.433..528D}.  With \textsl{Swift} BAT, \citet{2012MNRAS.419.2695R} constrained the eclipse half-angle to be 0.17$\pm$0.05 of the orbital period, corresponding to a duration of 0.6$\pm$0.2\,d (55$\pm$16\,ks).  The duration of the eclipse was found to be $\sim$0.75\,d in the archival data set from \textsl{INTEGRAL}/IBIS, covering the time period MJD\,52650--55469 \citep{2013MNRAS.433..528D}.  While the estimate in \citet{2013MNRAS.433..528D} is significantly larger than the constraints from \citet{2012MNRAS.419.2695R}, a lower limit of $\sim$0.583\,d was found in a combined study with \textsl{INTEGRAL}/IBIS and \textsl{XMM-Newton} \citep{2013MNRAS.433..528D}.

%figure 7
\begin{figure}[ht]
\centerline{\includegraphics[angle=-90,width=3in]{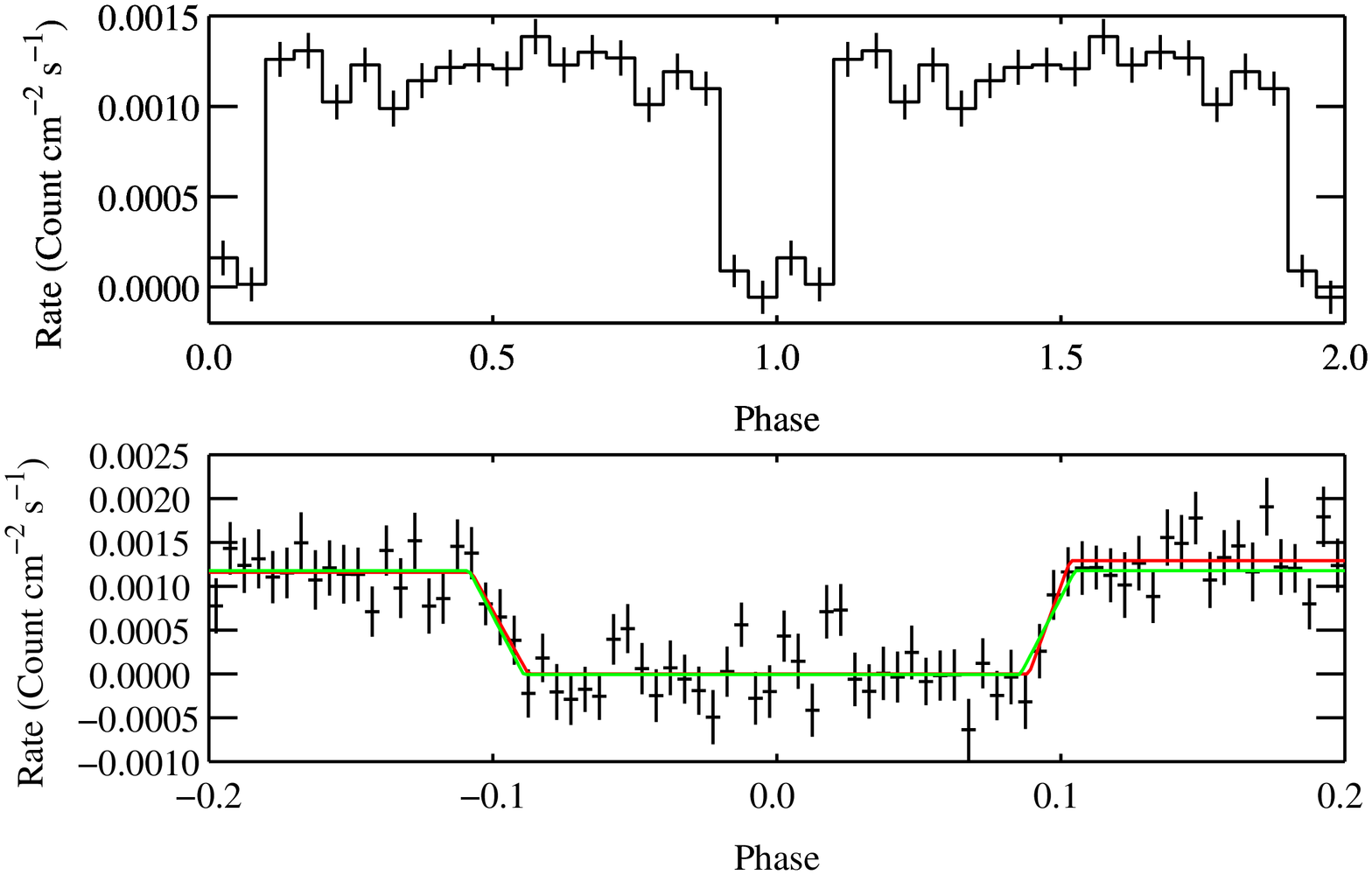}}
\figcaption[april22-2015_BAT_proper-folded-lightcurves_IGRJ16418-4532.ps]{
\textsl{Swift}-BAT light curve of IGR J16418-4532 {\mybf in the 15--50\,keV band} folded on the orbital period (top) using 20 bins.  T0 is defined at BMJD\,55087.721, corresponding to mid-eclipse.  A detailed folded light curve with 80 bins (bottom) is fit with {\mybf both a symmetric ``step and ramp" function (green) and asymmetric ``step and ramp" function (red)}, which model the eclipse.  The symmetric ``step and ramp" function was shifted accordingly.
\label{J16418 Folded Half Angle}
}
\end{figure}

\begin{figure}[ht]
\centerline{\includegraphics[width=3in]{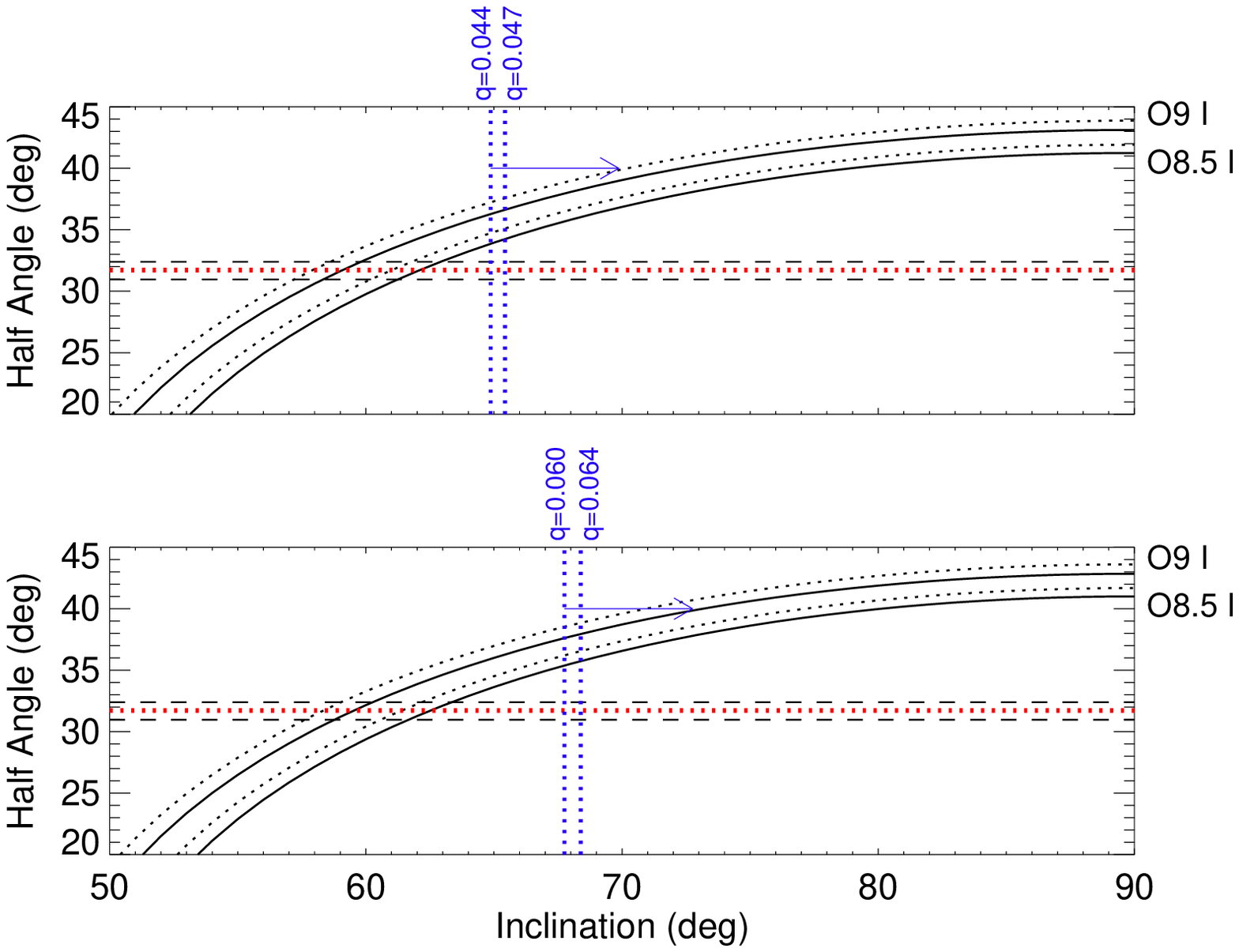}}
\figcaption[3-30-2015_IGRJ16418-4532_weighted_eclipse_param.eps]{
The black curves show the predicted eclipse half angle of IGR J16418-4532 as a function of inclination angle for stars with the indicated spectral types.  The red and black dashed lines indicate the eclipse half angle and estimated error as measured by \textsl{Swift} BAT.  We assume a neutron star mass of 1.4\,$M_\sun$ (top) and of mass 1.9\,$M_\sun$ (bottom) and typical masses and radii for the assumed companion spectral type (see Table~\ref{J16418 Primary Parameters}).  The blue vertical dashed lines indicate the lower limit of the inclination angle.  Inclinations to the left of these correspond to stars that overfill the Roche-lobe.
\label{J16418 Eclipse Half Angle}
}
\end{figure}

\begin{figure}[ht]
\centerline{\includegraphics[width=3in]{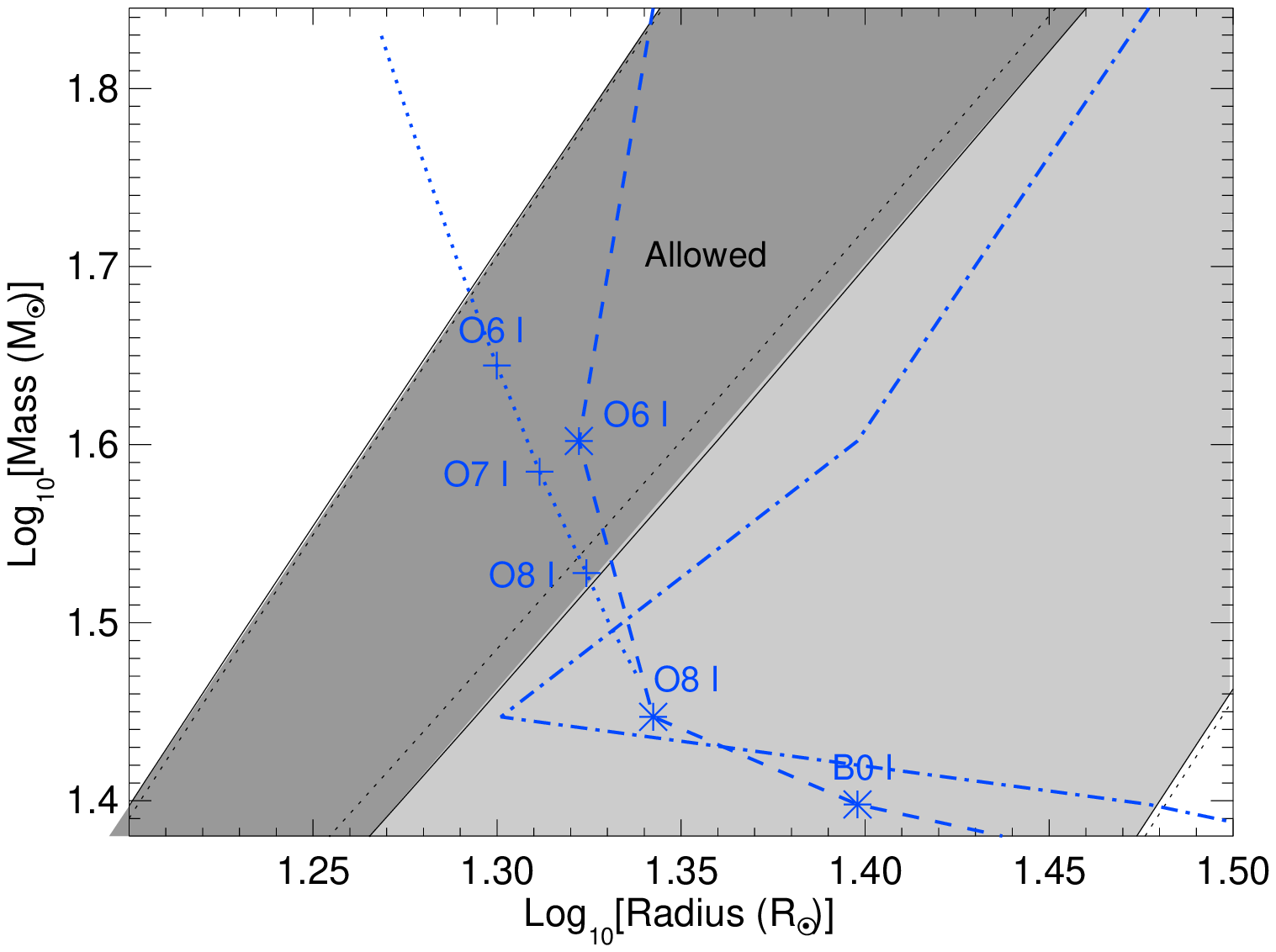}}
\figcaption[march30-2015_IGRJ16418-4532_mass-constraints.eps]{
Log-log plot of stellar mass as a function of stellar radius for IGR J16418-4532.  The dark shaded region indicates the allowed spectral types that satisfy the constraints imposed by both the eclipse duration and Roche-lobe size.  The light shaded region only indicates spectral types that satisfy the observed eclipse duration.  Stellar masses and radii are reported in Table~\ref{J16418 Primary Parameters}.  The dashed, dotted and dash-dotted lines indicate stellar masses and radii derived from \citet{2006ima..book.....C}, \citet{2005A&A...436.1049M} and \citet{2000asqu.book.....A}, respectively.
\label{J16418 Mass Radius Plot}
}
\end{figure}

We derive a 3.73863$\pm$0.00015\,d orbital period for IGR J16418-4532 using the fundamental peak in the power spectrum, while the first harmonic yields 3.73882$\pm$0.00011\,d.  Using an $O-C$ analysis (see {\mybf Figures~\ref{O-C Residuals}--\ref{O-C Historic}}), we refine this to {\mybf 3.73881$\pm$0.00002\,d}.  {\mybf Using the quadratic orbital change function (see Equation~\ref{Orbital Change Function}), we find the orbital period derivative to be 0.7$\pm$1.0$\times$10$^{-7}$\,d d$^{-1}$, which is consistent with zero.}  Folding the light curve on our refined orbital period (see Figure~\ref{J16418 Folded Half Angle}), we calculate the duration of the observed eclipse to be {\mybf 57$\pm$1\,ks (0.66$\pm$0.01\,d)}.  This yields an eclipse half-angle of {\mybf 31.7$^{+0.7}_{-0.8}$$\degr$} (see Table~\ref{Asymmetric Step and Ramp Model}).  {\mybf We note that we find the eclipse half-angle to be 31.5$\pm$0.6$\degr$ assuming a symmetric step-and-ramp function.}  We find the eclipse properties to be consistent with the results from \citet{2012MNRAS.419.2695R} and \citet{2013MNRAS.433..528D}.  Under the assumption that the mass and radius of the proposed mass donor are 31.54\,$M_\sun$ and 21.41\,$R_\sun$ \citep{2005A&A...436.1049M}, the duration of the observed eclipse is consistent with the proposed mass donor, where we find the orbital inclination to be {\mybf 60--63$\degr$} for an O8.5 I spectral type (see Figures~\ref{J16418 Eclipse Half Angle} and~\ref{J16418 Mass Radius Plot}).

%figure 8

%figure 9

% Table 7
\begin{deluxetable}{ccccccccc}
\tablecolumns{9}
\tablewidth{0pc}
\tablecaption{Physical Parameters for Previously Proposed Mass Donors for IGR J16418-4532}
\tablehead{
\colhead{Spectral Type} & \colhead{$M/M_\sun$} & \colhead{$q$$^a$} & \colhead{$R/R_\sun$} & \colhead{$R_{\rm L}$$/R_\sun$$^b$} & \colhead{$i$$\degr$}} 
\startdata
\textsl{O8.5 I} & \textsl{31.54} & 0.044 & \textsl{21.41$^c$} & {\mybf 20.71$^c$} & {\mybf 61--63} \\
\textsl{O8.5 I} & \textsl{33.90} & 0.041 & \textsl{22.20$^c$} & {\mybf 21.35$^c$} & {\mybf 60--62} \\
\textsl{O9 I} & \textsl{29.63} & 0.047 & \textsl{21.76$^c$} & {\mybf 20.17$^c$} & {\mybf 58--60} \\
\textsl{O9 I} & \textsl{31.95} & 0.044 & \textsl{22.60$^c$} & {\mybf 20.82$^c$} & {\mybf 57--59} \\
\enddata
\tablecomments{The values in italics are obtained from \citet{2005A&A...436.1049M}. \\*
$^a$ The mass ratio, $q$, is defined as $M_{\rm x}/M_{\rm c}$ where $M_{\rm x}$ is the compact object and $M_{\rm c}$ is the donor star. \\*
$^b$ The definition for the Roche lobe, $R_{\rm L}$, as given in \citet{1983ApJ...268..368E}, assuming $M_{\rm NS}$ is 1.4\,$M_\sun$. \\*
$^c$ These spectral types significantly overfill the Roche-lobe and are therefore excluded from our analysis.
}
\label{J16418 Primary Parameters}
\end{deluxetable}

Since the X-ray luminosity is lower than what would be expected for Roche-lobe overflow \citep{2004RMxAC..21..128K}, we can attach an additional constraint assuming that the mass donor underfills the Roche-lobe radius (see Figure~\ref{J16418 Mass Radius Plot}).  We note that while the derived masses and radii for the spectral types from \citet{2005A&A...436.1049M} satisfy the eclipse half-angle, the assumed radius would be larger than the Roche-lobe radius \citep{1983ApJ...268..368E}.  Therefore, an O8.5 I spectral type must be excluded. 

\subsection{IGR J16479-4514}

IGR J16479-4514 is an intermediate SFXT, which has been proposed to host either an O8.5 I \citep{2008A&A...484..783C,2008A&A...484..801R} or a O9.5 Iab \citep{2008A&A...486..911N} mass donor.  First discovered by the \textsl{INTEGRAL} satellite in 2003 August \citep{2003ATel..176....1M}, the fluxes in the 18--25\,keV and 25--50\,keV energy bands were found to be $\sim$12\,mCrab and $\sim$8\,mCrab, respectively.  The flux was later shown to increase by a factor of $\sim$2 on 2003 August 10 \citep{2003ATel..176....1M}. Using \textsl{Swift} BAT data, \citet{2009MNRAS.397L..11J} found the presence of a 3.319$\pm$0.001\,day modulation, which was interpreted as the orbital period.  A 3.3193$\pm$0.0005\,day modulation was independently found by \citet{2009MNRAS.399.2021R} also using BAT.  {\mybf \citet{2013ApJ...778...45C} found the orbital period to be 3.3199$\pm$0.0005\,d, which is consistent with the results from \citet{2009MNRAS.397L..11J} and \citet{2009MNRAS.399.2021R}.}  The presence of a 11.880$\pm$0.002\,d superorbital period was found by \citet{2013ApJ...778...45C} using BAT.  \citet{2013ATel.5131....1D} reported a 11.891$\pm$0.002\,d superorbital period using \textsl{INTEGRAL}, confirming the result.  No pulse period has been identified.

The BAT light curves folded on the orbital period revealed a sharp dip, which was interpreted as an eclipse with a proposed duration of $\sim$52\,ks \citep{2009MNRAS.397L..11J}.  This confirmed an earlier \textsl{XMM-Newton} observation where a decay from a higher to lower flux state was interpreted as the ingress of an eclipse \citep{2009AIPC.1126..319B}.  A 2012 \textsl{Suzaku} observation covered $\sim$80 percent of the orbital cycle of IGR J16479-4514, where temporal and spectral properties were analyzed during eclipse and out-of-eclipse \citep{2013MNRAS.429.2763S}.  Since the ingress of the eclipse was not covered in \textsl{Suzaku} observation, the exact duration of the eclipse could only be constrained to 46--143\,ks (0.53--1.66\,d) \citep{2013MNRAS.429.2763S}.

%figure 10
\begin{figure}[ht]
\centerline{\includegraphics[width=3in]{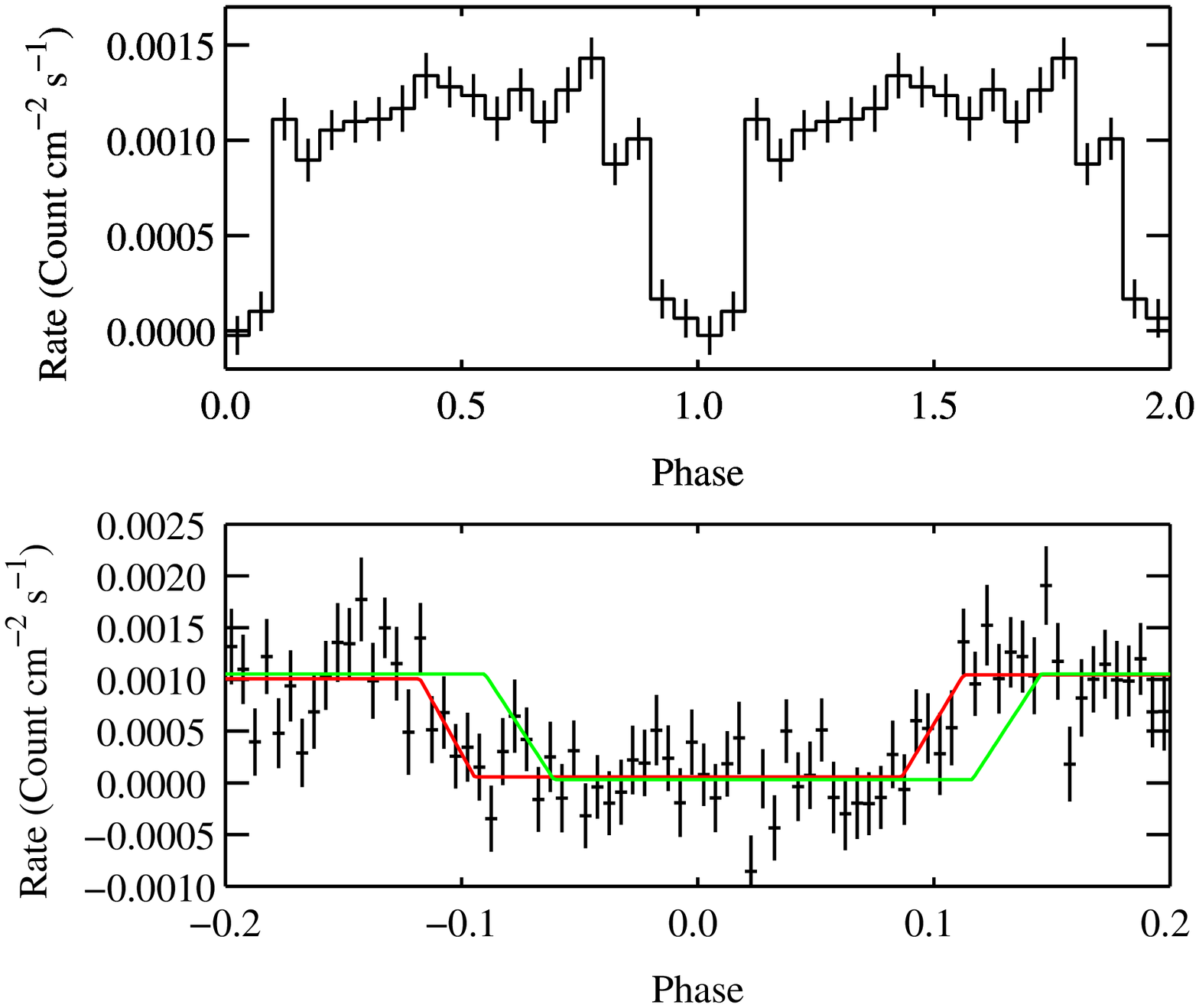}}
\figcaption[april29-2015_BAT_proper-folded-lightcurves_IGRJ16479-4514.ps]{
\textsl{Swift}-BAT light curve of IGR J16479-4514 {\mybf in the 15--50\,keV band} folded on the orbital period (top) using 20 bins.  T0 is defined at BMJD\,55081.57, corresponding to mid-eclipse.  A detailed folded folded light curve with 80 bins (bottom) is fit with {\mybf both a symmetric ``step and ramp" function (green) and asymmetric ``step and ramp" function (red)}, which models the eclipse.  The symmetric ``step and ramp" function was shifted accordingly.
\label{J16479 Folded Half Angle}
}
\end{figure}

\begin{figure}[ht]
\centerline{\includegraphics[width=3in]{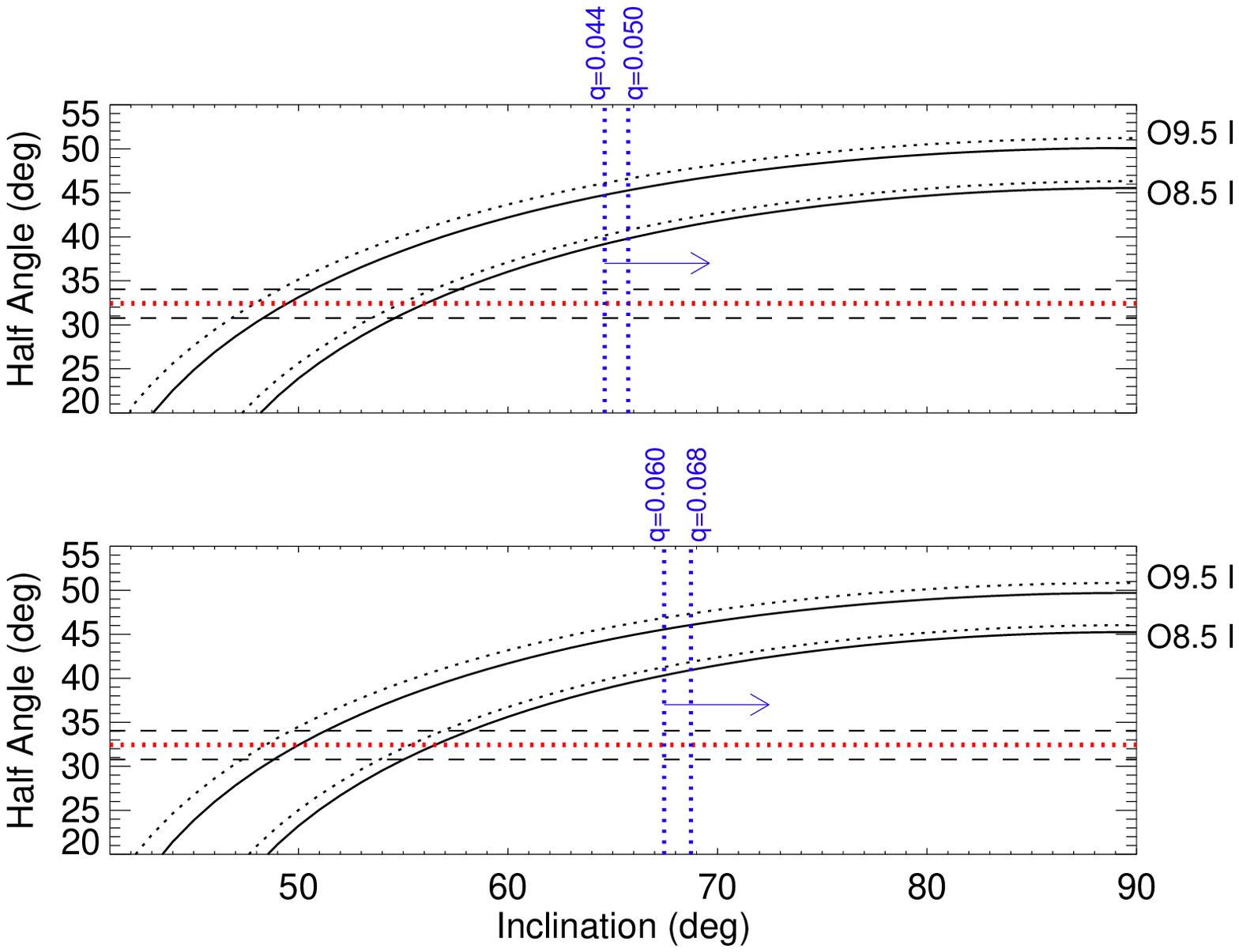}}
\figcaption[4-29-2015_IGRJ16479-4514_corrected_eclipse_param.eps]{
The black curves show the predicted eclipse half angle of IGR J16479-4514 as a function of inclination angle for stars with the indicated spectral types.  The red and black dashed lines indicate the eclipse half angle and estimated error as measured by \textsl{Swift} BAT.  We assume a neutron star mass of 1.4\,$M_\sun$ (top) and of mass 1.9\,$M_\sun$ (bottom) and typical masses and radii for the assumed companion spectral type (see Table~\ref{J16479 Primary Parameters}).  The blue vertical dashed lines indicate the lower limit of the inclination angle.  Inclinations to the left of these correspond to stars that overfill the Roche-lobe.
\label{J16479 Eclipse Half Angle}
}
\end{figure}

\begin{figure}[ht]
\centerline{\includegraphics[width=3in]{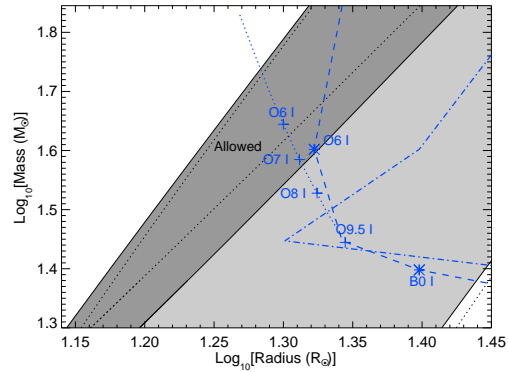}}
\figcaption[april29-2015_IGRJ16479-4514_mass-constraints.eps]{
Log-log plot of stellar masses as a function of stellar radius for IGR J16479-4514.  The light shaded region indicates the allowed spectral types that satisfy the observed eclipse.  The dark shaded region indicates the spectral types where both the eclipse duration and Roche-lobe radius constraints are satisfied.  Stellar masses and radii are reported in Table~\ref{J16479 Primary Parameters}.  The dashed, dotted and dash-dotted lines indicate stellar masses and radii derived from \citet{2006ima..book.....C}, \citet{2005A&A...436.1049M} and \citet{2000asqu.book.....A}, respectively.
\label{J16479 Mass Radius Plot}
}
\end{figure}

Using a DFT, we derive the orbital period of IGR J16479-4514 to be {\mybf 3.31998$\pm$0.00014\,d}.  We refine this to {\mybf 3.31961$\pm$0.00004\,d} using an $O-C$ analysis (see {\mybf Figures~\ref{O-C Residuals}--\ref{O-C Historic}}) and fold the light curve on our refined orbital period to calculate the eclipse half-angle (see Figure~\ref{J16479 Folded Half Angle}).  {\mybf Using the quadratic orbital change function (see Equation~\ref{Orbital Change Function}), we find the orbital period derivative to be 3$\pm$2$\times$10$^{-7}$\,d d$^{-1}$, which is consistent with zero.}  We calculate the duration of the observed eclipse to be {\mybf 52$\pm$3\,ks (0.60$\pm$0.03\,d)}, which is consistent with results from \citet{2009MNRAS.397L..11J}.  This yields an eclipse half-angle of {\mybf 31.9$^{+0.9}_{-1.3}$}$\degr$ (see Table~\ref{Step and Ramp Model}).  Using values from \citet{2005A&A...436.1049M} for the masses and radii of the proposed spectral type of the mass donor, the duration of the observed eclipse is consistent with the proposed mass donor (see Figures~\ref{J16479 Eclipse Half Angle}), where we find the orbital inclination to be {\mybf 54}--58$\degr$ or 47--51$\degr$ for a O8.5 I or O9.5 Iab spectral type, respectively (see Table~\ref{J16479 Primary Parameters}).  While the previously preposed spectral types satisfy the eclipse half-angle, we note that the radius of the proposed spectral type is larger than the Roche-lobe \citep{1983ApJ...268..368E}.  Therefore, the previously proposed O8.5 I or O9.5 Iab spectral types must be excluded (see Figure~\ref{J16479 Mass Radius Plot}).  \citet{2013MNRAS.429.2763S} proposed a mass donor with mass $\sim$35\,$M_\sun$ and radius $\sim$20\,$R_\sun$, which could satisfy the constraints imposed by the Roche-lobe radius.

%figure 11

%figure 12

% Table 8
\begin{deluxetable}{ccccccccc}
\tablecolumns{9}
\tablewidth{0pc}
\tablecaption{Physical Parameters for Previously Proposed Mass Donors for IGR J16479-4514}
\tablehead{
\colhead{Spectral Type} & \colhead{$M/M_\sun$} & \colhead{$q$$^a$} & \colhead{$R/R_\sun$} & \colhead{$R_{\rm L}$$/R_\sun$$^b$}  & \colhead{$i$$\degr$}} 
\startdata
\textsl{O8.5 I} & \textsl{31.54} & 0.044 & \textsl{21.41$^c$} & {\mybf 19.13$^c$} & {\mybf 55--58} \\
\textsl{O8.5 I} & \textsl{33.90} & 0.041 & \textsl{22.20$^c$} & {\mybf 19.73$^c$} & {\mybf 54--56} \\
\textsl{O9.5 I} & \textsl{27.83} & 0.050 & \textsl{22.11$^c$} & {\mybf 18.14$^c$} & {\mybf 48--51} \\
\textsl{O9.5 I} & \textsl{30.41} & 0.046 & \textsl{23.11$^c$} & {\mybf 18.84$^c$} & {\mybf 47--49} \\
\enddata
\tablecomments{The values in italics are obtained from \citet{2005A&A...436.1049M}. \\*
$^a$ The mass ratio, $q$, is defined as $M_{\rm x}/M_{\rm c}$ where $M_{\rm x}$ is the compact object and $M_{\rm c}$ is the donor star. \\*
$^b$ The definition for the Roche lobe, $R_{\rm L}$, as given in \citet{1983ApJ...268..368E}, assuming $M_{\rm NS}$ is 1.4\,$M_\sun$. \\*
$^c$ These spectral types significantly overfill the Roche-lobe and are therefore excluded from our analysis.
}
\label{J16479 Primary Parameters}
\end{deluxetable}

\subsection{IGR J18027-2016 (=SAX J1802.7-2017)}
\label{IGR J18027-2016 Results}

IGR J18027-2016 (=SAX J1802.7-2017) is an SGXB, which has been proposed to host either a B1 Ib \citep{2010A&A...510A..61T} or B0-B1 I \citep{2011A&A...532A.124M} mass donor.  First detected with \textsl{BeppoSAX} in 2001 September \citep{2003ApJ...596L..63A}, the average flux in the 0.1--10\,keV energy band was found to be 3.6$\times$10$^{-11}$\,ergs cm$^{-2}$ s$^{-1}$.  Pulse-timing analysis suggested a $\sim$4.6\,day orbital period \citep{2003ApJ...596L..63A}, which was later refined to 4.5696$\pm$0.0009\,days using \textsl{INTEGRAL} \citep{2005A&A...439..255H}.  {\mybf Combining the mid-eclipse times derived in \citet{2003ApJ...596L..63A} and \citet{2005A&A...439..255H} with later observations with \textsl{Swift} BAT and \textsl{INTEGRAL} in an $O-C$ analysis, the orbital period was further refined to 4.5693$\pm$0.0004\,d \citep{2009RAA.....9.1303J}.  Fitting a quadratic model to the derived mid-eclipse times, a period derivative of (3.9$\pm$1.2)$\times$10$^{-7}$\,d\,d$^{-1}$ was found \citep{2009RAA.....9.1303J}.  Using \textsl{INTEGRAL}, \citet{2015A&A...577A.130F} recently refined the orbital period and period derivative to 4.5697$\pm$0.0001\,d and (2.1$\pm$3.6)$\times$10$^{-7}$\,d\,d$^{-1}$, respectively.} \citet{2003ApJ...596L..63A} found the presence of a $\sim$139.6\,s modulation {\mybf using \textsl{BeppoSAX}}, which they interpeted as the neutron star rotation period.  \citet{2005A&A...439..255H} found the pulse period using \textsl{XMM-Newton} to be 139.61$\pm$0.04\,s, which is consistent with an earlier result from \citet{2003ApJ...596L..63A}.  However, no evidence for superorbital modulation was found \citep{2013ApJ...778...45C}.

%figure 13
\begin{figure}[ht]
\centerline{\includegraphics[angle=-90,width=3in]{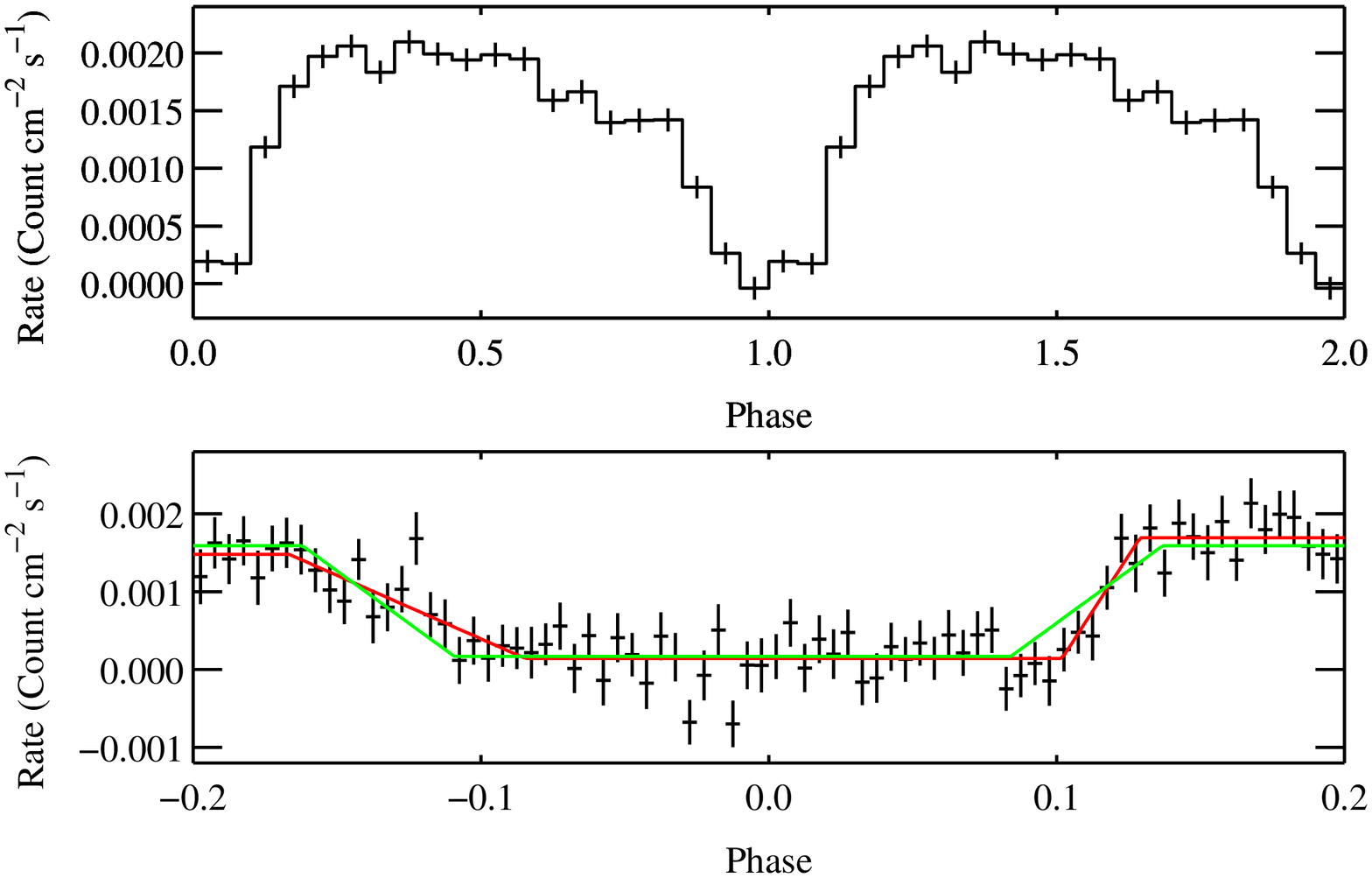}}
\figcaption[4-22-2015_BAT_proper-folded-lightcurves_IGRJ18027-2016.ps]{
\textsl{Swift}-BAT light curve of IGR J18027-2016 {\mybf in the 15--50\,keV band} folded on the orbital period (top) using 20 bins.  T0 is defined at BMJD\,55083.82$\pm$0.01, corresponding to mid-eclipse.  A detailed folded light curve with 80 bins (bottom) is fit with {\mybf both a symmetric ``step and ramp" function (green) and asymmetric ``step and ramp" function (red)}, which model the eclipse.  The symmetric ``step and ramp" function was shifted accordingly.
\label{J18027 Folded Half Angle}
}
\end{figure}

\citet{2003ApJ...596L..63A} derived the epoch of the NS superior conjuction to occur during the first 50\,ks of an observation where IGR J18027-2016 was undetected.  Since the epoch of superior conjuction corresponds to a time where the mid-eclipse is expected to occur, \citet{2003ApJ...596L..63A} suggested the presence of a possible eclipse with a half-duration of 0.47$\pm$0.10\,days.  \citet{2005A&A...439..255H} {\mybf and \citet{2009RAA.....9.1303J}} later confirmed this result where the eclispe half-angle was found to be 0.61$\pm$0.08\,radians (34.9$\pm$4.6\,$\degr$) {\mybf and $\sim$0.604\,radians ($\sim$34.6\,$\degr$), respectively}.  \citet{2015A&A...577A.130F} recently refined the eclipse half-angle to 31$\pm$2\,$\degr$ using \textsl{INTEGRAL}.

Pulse-timing and radial velocity curves have helped place constraints on the physical properties on both the donor star as well as the compact object.  Using a pulse-timing analysis with \textsl{BeppoSAX} and \textsl{XMM-Newton}, the projected semi-major axis of the neutron star was found to be 68$\pm$1\,lt-s \citep{2005A&A...439..255H}.  Assuming a 1.4\,$M_\sun$ neutron star, \citet{2005A&A...439..255H} constrained the mass and radius of the mass donor star to 18.8--29.3\,$M_\sun$ and 14.7--23.4\,$R_\sun$, respectively.  {\mybf Using \textsl{Swift} BAT and \textsl{INTEGRAL}, \citet{2009RAA.....9.1303J} further constrained the radius of the donor star to 16.4--24.7\,$R_\sun$.} The mass donor was observed between 2010 May 26 and 2010 September 8 (MJD\,55342--55447) with the \textsl{European Southern Observatory Very Large Telescope} \citep[\textsl{ESO VLT},][]{2011A&A...532A.124M}.  The {\mybf semi-amplitude of the} radial velocity of the mass donor, $K_{\rm O}$, was found to be 23.8$\pm$3.1\,km s$^{-1}$ \citep{2011A&A...532A.124M}.  Since the projected semi-major axis of the neutron star can be expressed in terms of a radial velocity {\mybf semi-amplitude}, $K_{\rm X}$, the ratio between the masses of the compact object and the mass donor can be calculated according to Equation {\mybf 2} in {\mybf \citet{1999MNRAS.307..357A}}.  \citet{2011A&A...532A.124M} found the mass ratio, q, to be 0.07$\pm$0.01.  Using the mass ratio and the eclipse half-angle measured in \citet{2005A&A...439..255H}, the mass and radius of the donor star were refined to values between 18.6$\pm$0.8\,$M_\sun$ and 16.8$\pm$1.5\,$R_\sun$ at edge-on inclinations to 21.8$\pm$2.4\,$M_\sun$ and 19.8$\pm$0.7\,$R_\sun$ where the donor star fills the Roche-lobe \citep{2011A&A...532A.124M}.  The mass of the compact object was constrained to be between 1.36$\pm$0.21 and 1.58$\pm$0.27\,$M_\sun$ in the two limits. The large error on the estimate of the eclipse half-angle from \citet{2005A&A...439..255H} contributes significantly to the uncertainties on these measurements.  Using a similar analysis with \textsl{INTEGRAL}, \citet{2015A&A...577A.130F} constrained the mass of the neutron star to 1.6$\pm$0.3\,$M_\sun$.

%figure 14
\begin{figure}[ht]
\centerline{\includegraphics[width=3in]{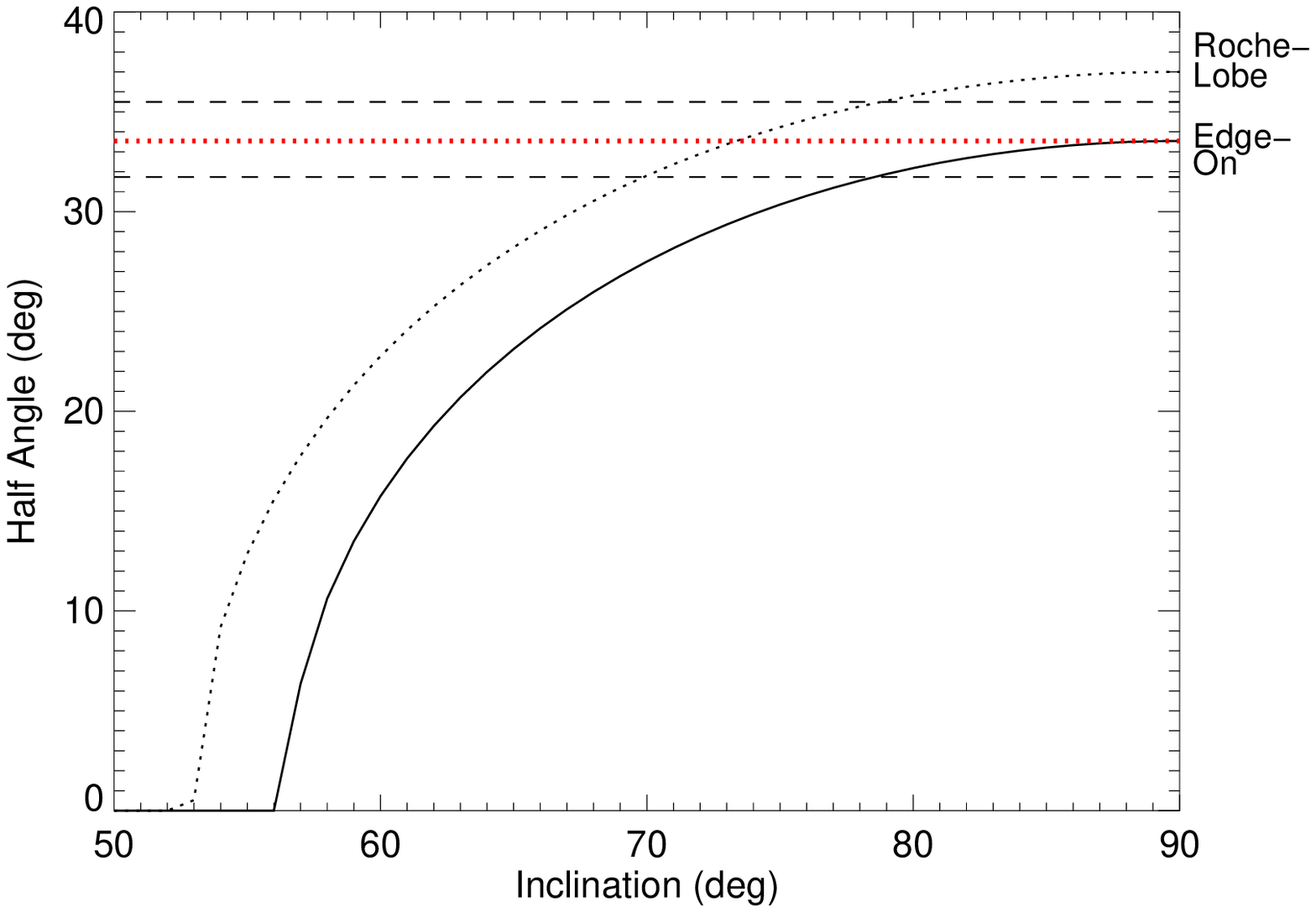}}
\figcaption[4-13-2015_IGRJ18027-2016_weighted_eclipse_param.eps]{
The black curves show the predicted eclipse half angle of IGR J18027-2016 as a function of inclination angle for stars with the indicated spectral types.  The red and black dashed lines indicate the eclipse half angle and estimated error as measured by \textsl{Swift} BAT.
\label{J18027 Eclipse Half Angle}
}
\end{figure}

%figure 15
\begin{figure}[ht]
\centerline{\includegraphics[width=3in]{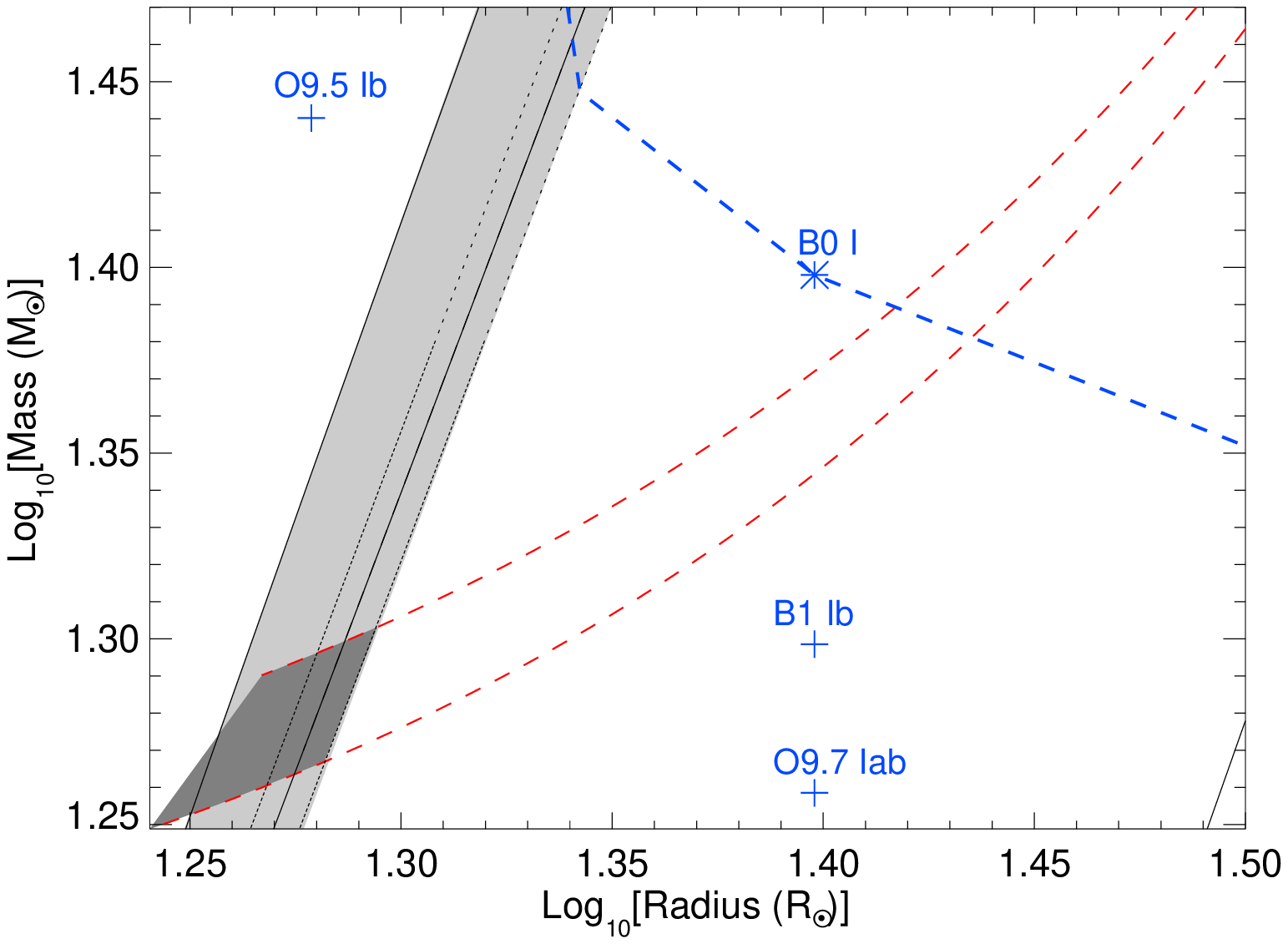}}
\figcaption[dec4-2014_IGRJ18027-2016_mass-constraints.eps]{
Log-log plots of stellar masses as a function of stellar radii for IGR J18027-2016.  The shaded region indicates the allowed spectral types to satisfy the observed eclipse duration and pulse-timing constraints.  Stellar masses and radii are reported in Table~\ref{J18027 Primary Parameters}.  Stars and crosses indicate the spectral types according to \citet{2006ima..book.....C} and \citet{2007A&A...463.1093L}, respectively.
\label{J18027 Mass Radius Plot}
}
\end{figure}

% Table 9
\begin{deluxetable}{cc}
\tablecolumns{3}
\tablewidth{0pc}
\tablecaption{System Parameters for IGR J18027-2016}
\tablehead{
\colhead{Parameter} & \colhead{Value}}
\startdata
$P_{\rm orb}$$^a$ & {\mybf 4.56993$\pm$0.00003\,d} \\
$P_{\rm pulse}$$^b$ & 139.61$\pm$0.04\,s \\
$a_{\rm x}$ $sin$\textsl{i}$^b$ & 68$\pm$1\,lt-s \\
$K_{\rm o}$$^c$ & 23.8$\pm$3.1\,km s$^{-1}$ \\
$T_{\rm mid}$ & {\mybf BMJD\,55083.82$\pm$0.01} \\
$f(M)$$^b$ & 16$\pm$1\,$M_\sun$ \\
$q$$^c$ & 0.07$\pm$0.01 \\
$\Theta_{\rm e}$ & {\mybf 34$\pm$2}$\degr$ \\
\tableline
$M_{\rm donor}$ & {\mybf 18.6$\pm$0.9}--{\mybf 19.4$\pm$0.9}\,$M_\sun$ \\
$R_{\rm donor}$ & {\mybf 17.4$\pm$0.9}--{\mybf 19.5$^{+0.8}_{-0.7}$}\,$R_\sun$ \\
$a$ & 31.4--33.2\,$R_\sun$ \\
$M_{\rm NS}$ & {\mybf 1.37$\pm$0.19}--{\mybf 1.43$\pm$0.20}\,$M_\sun$ \\
$i$ & {\mybf 73.3}--90$\degr$ \\
\enddata
\tablecomments{
$^a$ The orbital period is refined using an $O-C$ analysis. \\
$^b$ The pulse period, projected semi-major axis and mass function are given in \citet{2005A&A...439..255H}. \\*
$^c$ The {\mybf semi-amplitude of the} radial velocity of the mass donor and mass ratio are given in \citet{2011A&A...532A.124M}.}
%}
\label{J18027 Primary Parameters}
\end{deluxetable}

We derive a {\mybf 4.57022$\pm$0.00013\,d} orbital period for IGR J18027-2016 using a DFT, which is consistent with the results from \citet{2005A&A...439..255H}.  Using an $O-C$ analysis (see {\mybf Figures~\ref{O-C Residuals}--\ref{O-C Historic}}), we refine this to {\mybf 4.56993$\pm$0.00003\,d}.  {\mybf Using the quadratic orbital change function (see Equation~\ref{Orbital Change Function}), we find the orbital period derivative to be 0.2$\pm$1.1$\times$10$^{-7}$\,d d$^{-1}$, which is consistent with zero.}  Folding the light curve on our refined orbital period(see Figure~\ref{J18027 Folded Half Angle}), we {\mybf measure} the duration of the observed eclipse to be {\mybf 74$\pm$4\,ks (0.85$\pm$0.05\,d)}.  We find the eclipse half-angle to be {\mybf 34$\pm$2}$\degr$ (see Table~\ref{Step and Ramp Model}).  Since the X-ray luminosity of IGR J18027-2016 is modest, we again attach the constraint that the donor star underfills the Roche-lobe (see Figures~\ref{J18027 Eclipse Half Angle} and~\ref{J18027 Mass Radius Plot}).  This constrains the mass and radius of the donor star as well as the mass of the neutron star (see Figure~\ref{J18027 Mass Radius Plot}).  Using the eclipse half-angle and the expression for the mass donor when the mass ratio is known \citep[Eq.4;][]{2011A&A...532A.124M}, we calculate the donor star mass and radius as well as mass of the compact object.  We find the mass and radius of the donor star to be {\mybf 18.6$\pm$0.9}\,$M_\sun$ and {\mybf 17.4$\pm$0.9}\,$R_\sun$ and {\mybf 19.4$\pm$0.9}\,$M_\sun$ and {\mybf 19.5$^{+0.8}_{-0.7}$}\,$R_\sun$ in the two limits (see Table~\ref{J18027 Primary Parameters}).  In the allowed limits, we constrain the mass of the neutron star to between {\mybf 1.37$\pm$0.19}\,$M_\sun$ and {\mybf 1.43$\pm$0.20}\,$M_\sun$ (see Table~\ref{J18027 Primary Parameters}). While our results are in agreement with calculations in \citet{2011A&A...532A.124M}, we note that the error estimate is {\mybf only marginally} improved in our analysis.  {\mybf The driving factor on the error estimate of the neutron star mass is the uncertainty of $\sim$13$\%$ on the {\mybf semi-amplitude of the} radial velocity of the donor star as reported in \citet{2011A&A...532A.124M}.}

\subsection{XTE J1855-026}
\label{XTE J1855-026 Results}

XTE J1855-026 is an SGXB discovered during \textsl{RXTE} scans along the Galactic plane \citep{1999ApJ...517..956C}.  Through 11 scanning observations of the Scutum arm, the 2--10\,keV flux of XTE J1855-026 varied from an upper limit of 10\,counts s$^{-1}$ to 136$\pm$15\,counts s$^{-1}$ \citep{1999ApJ...517..956C}.  Using the \textsl{RXTE} Proportional Counter Array (PCA), \citet{1999ApJ...517..956C} found a 361.1$\pm$0.4\,s modulation, which was interpreted as the neutron star rotation period.  \citet{2002ApJ...577..923C} later refined the pulse period to 360.741$\pm$0.002\,s with the PCA.  An analysis using the \textsl{RXTE} All-sky Monitor (ASM) revealed the presence of a 6.0724$\pm$0.0009\,d modulation, which is interpreted as the orbital period \citep{2002ApJ...577..923C}.  {\mybf Using an $O-C$ analysis with \textsl{INTEGRAL}, \citet{2015A&A...577A.130F} recently refined this to 6.07415$\pm$0.00008\,d.  No significant orbital period derivative was found using the quadratic orbital change function \citep{2015A&A...577A.130F}.}  This places XTE J1855-026 in the wind-fed supergiant region of the Corbet diagram \citep{1986MNRAS.220.1047C}.  \citet{2002ApJ...577..923C} constrained the eccentricity of XTE J1855-026 to e$\leq$0.04 using pulse-timing analysis.  The projected semi-major axis of the neutron star was found to be 82.8$\pm$0.8\,lt-s for a circular orbit and 80.5$\pm$1.4\,lt-s for the eccentric solution \citep{2002ApJ...577..923C}.  In this section, we consider the scenario where the orbit is circular.  We calculate constraints on the mass and radius of the donor star in the more complicated scenario where the orbit is considered to have a modest eccentricity in Section~\ref{Constraints on eccentricity}.  Using optical and near-infrared spectroscopy obtained with the William Herschel Telescope (WHT), \citet{2008ATel.1876....1N} found the mass donor to be a supergiant with spectral type B0 Iaep.

The light curves folded on the orbital period reveal a sharp dip, which was interpreted as an eclipse with a total phase duration of 0.198--0.262 \citep{2002ApJ...577..923C}.  The phase duration corresponded to an eclipse half-angle 36$\degr$$\leq$$\Theta_{\rm e}$$\leq$47$\degr$.  The eclipse duration was found to be 93$\pm$3\,ks (1.08$\pm$0.03\,d) in the archival \textsl{INTEGRAL} data set \citep{2015A&A...577A.130F}, corresponding to an eclipse half-angle of 32$\pm$1$\degr$.  This measurement is somewhat lower than result from \citet{2002ApJ...577..923C}.

{\mybf Optical radial velocity curves recently obtained with the \textsl{Isaac Newton Telescope (INT)}, the \textsl{Liverpool Telescope (LT)} and the \textsl{WHT} help place additional constraints on the components of the system \citep{2015arXiv150301087G}.  The {\mybf semi-amplitude of the} radial velocity of the donor star was found to be 26.8$\pm$8.2\,km s$^{-1}$.  Expressing the projected semi-major axis of the neutron star as a radial velocity {\mybf semi-amplitude}, \citet{2015arXiv150301087G} found the ratio between the masses of the components of the system to be 0.09$\pm$0.03 and notes that a large eccentricity of $\sim$0.4--0.5 was found in the optical orbital solutions.  This strongly contrasts with the X-ray orbital solution reported in \citet{2002ApJ...577..923C}, suggesting that caution must be taken in interpreting the optical orbital solutions.  \citet{2015arXiv150301087G} refined the spectral type of the mass donor to a BN0.2 Ia supergiant and found the mass and radius of the donor star to be 13$^{+19}_{-7}$\,$M_\sun$ and 27$^{+21}_{-10}$\,$R_\sun$, respectively.}

%figure 16
\begin{figure}[ht]
\centerline{\includegraphics[angle=-90,width=3in]{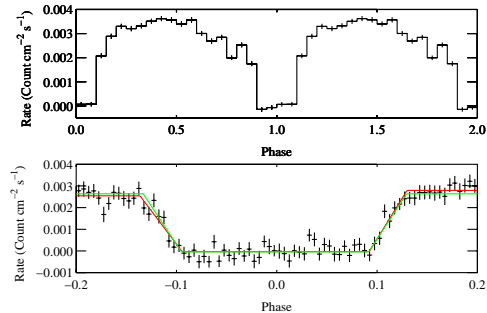}}
\figcaption[april22-2015_BAT_proper-folded-lightcurves_XTEJ1855-026.ps]{
\textsl{Swift}-BAT light curve of XTE J1855-026 {\mybf in the 15--50\,keV band} folded on the orbital period (top) using 20 bins.  T0 is defined as {\mybf MJD\,55079.0685}, corresponding to mid-eclipse.  A detailed folded folded light curve with 80 bins (bottom) is fit with {\mybf both a symmetric ``step and ramp" function (blue) and asymmetric ``step and ramp" function (red)}, which models the eclipse.
\label{J1855 Folded Half Angle}
}
\end{figure}

\begin{figure}[ht]
\centerline{\includegraphics[width=3in]{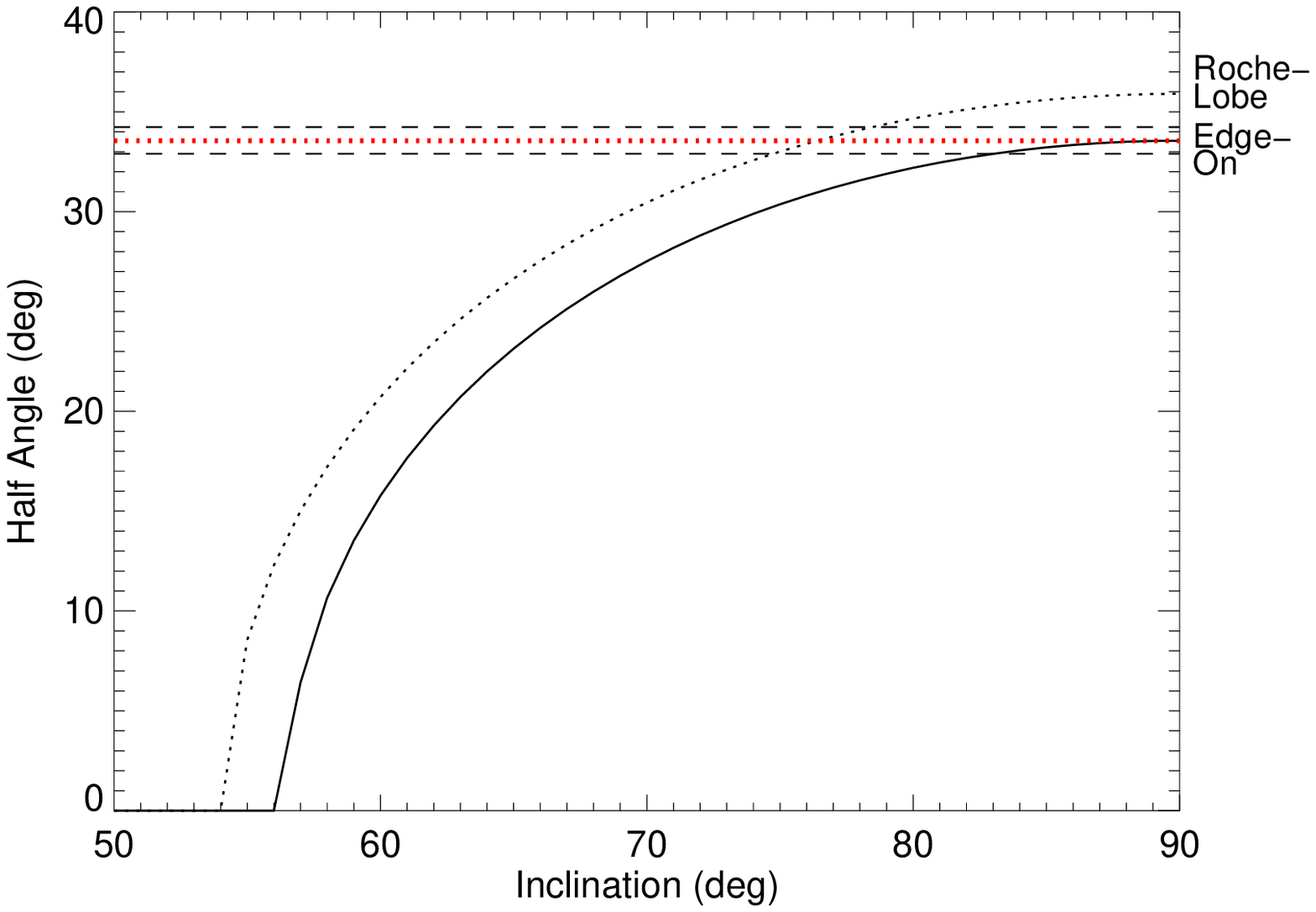}}
\figcaption[april1-2015_XTEJ1855-026_weighted_eclipse_param.eps]{
The black curves show the predicted eclipse half angle of XTE J1855-026 as a function of inclination angle for stars with the indicated spectral types.  The red and black dashed lines indicate the eclipse half angle and estimated error as measured by \textsl{Swift} BAT.
\label{J1855 Eclipse Half Angle}
}
\end{figure}

We derive a {\mybf 6.07411$\pm$0.00014\,d} orbital period for XTE J1855-026 using the fundamental peak in the power spectrum.  Using an $O-C$ analysis (see {\mybf Figures~\ref{O-C Residuals}--\ref{O-C Historic}}), we refine this to {\mybf 6.07413$\pm$0.00004\,d} and fold the light curve on the refined orbital period to calculate the eclipse half-angle (see Figures~\ref{J1855 Folded Half Angle} and ~\ref{J1855 Eclipse Half Angle}).  We calculate the duration of the observed eclipse to be {\mybf 98$\pm$2\,ks (1.13$\pm$0.02\,d)}, yielding an eclipse half-angle of {\mybf 33.6$\pm$0.7\,degrees} (see Table~\ref{J1855 Primary Parameters}).  This indicates that our derived eclipse duration is somewhat less than the result from \citet{2002ApJ...577..923C} and is consistent with the measurement in \citet{2015A&A...577A.130F}.{\mybf Using the quadratic orbital change function (see Equation~\ref{Orbital Change Function}), we find the orbital period derivative to be 0.0$\pm$0.5$\times$10$^{-7}$\,d d$^{-1}$, which is consistent with zero.}

\begin{figure}[ht]
\centerline{\includegraphics[width=3in]{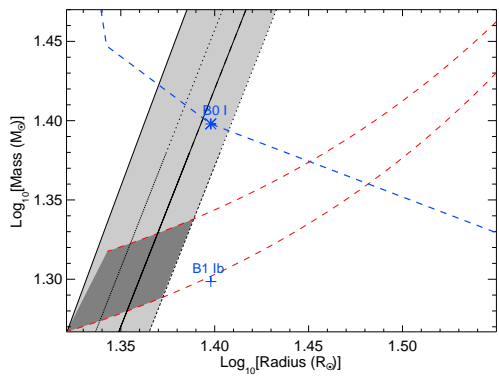}}
\figcaption[april23-2015_XTEJ1855-026_mass-constraints.eps]{
Log-log plots of stellar masses as a function of stellar radii for XTE J1855-026.  The shaded region indicates the allowed spectral types to satisfy the observed eclipse duration, pulse-timing and optical radial velocity constraints.  Stellar masses and radii are reported in Table~\ref{J1855 Primary Parameters}.  Stars and crosses indicate the spectral types according to \citet{2006ima..book.....C} and \citet{2007A&A...463.1093L}, respectively.
\label{J1855 Mass Radius Plot}
}
\end{figure}

{\mybf Since the upper limits of the stellar mass and radius are constrained by the orbital inclination where the Roche lobe is just filled, we again attach constraints on the mass and radius of the donor star as well as the mass of the compact object. We find the inclination where the donor star fills the Roche lobe to be 76.4$\degr$.  The stellar mass and radius of the donor star are found to be 19.6$\pm$1.1\,$M_\sun$ and 21.5$\pm$0.5\,$R_\sun$ and 20.2$\pm$1.2\,$M_\sun$ and 23.0$\pm$0.5\,$R_\sun$ in the two limits (see Table~\ref{J1855 Mass Radius Plot}).  We find the mass of the neutron star to be between 1.77$\pm$0.55\,$M_\sun$ and 1.82$\pm$0.57\,$M_\sun$ (see Table~\ref{J1855 Mass Radius Plot}), where the driving factor on the large error estimate is the uncertainty of $\sim$31$\%$ on the radial velocity {\mybf semi-amplitude} of the donor star as reported in \citet{2015arXiv150301087G}.}

%figure 17

%figure 18

% Table 10
\begin{deluxetable}{cc}
\tablecolumns{3}
\tablewidth{0pc}
\tablecaption{{\mybf System Parameters for XTE J1855-026}}
\tablehead{
\colhead{{\mybf Parameter}} & \colhead{{\mybf Value}}}
\startdata
{\mybf $P_{\rm orb}$$^a$} & {\mybf 6.07413$\pm$0.00004\,d} \\
{\mybf $P_{\rm pulse}$$^b$} & 360.741$\pm$0.002\,s \\
{\mybf $\dot{P}_{\rm pulse}$$^b$} & 1.5$\pm$3.6\,s s$^{-1}$$\times$10$^{-8}$ \\
{\mybf $a_{\rm x}$ $sin$\textsl{i}$^b$} & 82.8$\pm$0.8\,lt-s \\
{\mybf $K_{\rm o}$$^c$} & {\mybf 26.8$\pm$8.2\,km s$^{-1}$} \\
{\mybf $T_{\rm mid}$} & {\mybf MJD\,55079.07$\pm$0.01} \\
{\mybf $f(M)$$^b$} & 16.5$\pm$0.5\,$M_\sun$ \\
{\mybf $q$$^c$} & {\mybf 0.09$\pm$0.03} \\
{\mybf $\Theta_{\rm e}$} & {\mybf 33.6$\pm$0.7$\degr$} \\
\tableline
{\mybf $M_{\rm donor}$} & {\mybf 19.6$\pm$1.1--20.2$\pm$1.2\,$M_\sun$} \\
{\mybf $R_{\rm donor}$} & {\mybf 21.5$\pm$0.5--23.0$\pm$0.5\,$R_\sun$} \\
{\mybf $a$} & {\mybf 38.9--39.2\,$R_\sun$} \\
{\mybf $M_{\rm NS}$} & {\mybf 1.77$\pm$0.55\,$M_\sun$--1.82$\pm$0.57\,$M_\sun$} \\
{\mybf $i$} & {\mybf 76.4--90$\degr$} \\
\enddata
\tablecomments{
{\mybf $^a$ The orbital period is refined using an $O-C$ analysis.} \\
{\mybf $^b$ The pulse period, derivative of the pulse period, projected semi-major axis and mass function are given in \citet{2002ApJ...577..923C}.} \\*
{\mybf $^c$ The {\mybf semi-amplitude of the} radial velocity of the mass donor and mass ratio are given in  \citet{2015arXiv150301087G}.}}
%}
\label{J1855 Primary Parameters}
\end{deluxetable}

\section{Discussion}

We discuss our findings of IGR J16393-4643, IGR J16418-4532, IGR J16479-4514, IGR J18027-2016 and XTE J1855-055.  The radii for the previously proposed spectral types in IGR J16418-4532 and IGR J16479-4514 would significantly overfill the Roche-lobe, which suggests an earlier spectral type.  Below, we discuss in detail the nature of the mass donors in each system, mechanisms that could result in the residual emission observed in IGR J16393-4643 and comment on the nature of the eclipse profiles.

\subsection{What is the nature of the mass donors in each system?}

\subsubsection{IGR J16393-4643}

Our results show that stars with spectral type B0 V and B0-5 III satisfy the constraints imposed by the eclipse half-angle as well as the Roche-lobe (see Section~\ref{IGR J16393-4643 Results}).  While some supergiant stars such as a B0 I satisfy the eclipse half-angle, the required radius would be larger than the Roche-lobe.  We calculated the Roche-lobe radius for stars of spectral type B0 I to be {\mybf 18.5\,$R_\sun$}.  This is clearly smaller than the radii reported in \citet{2006ima..book.....C} and \citet{2000asqu.book.....A}, which are 25\,$R_\sun$ and 30\,$R_\sun$, respectively (see Table~\ref{J16393 Primary Parameters}).  Since the radius for this proposed supergiant is too large to satisfy the constraint imposed by the Roche-lobe, we suggest that if the donor star is a supergiant then it must be a slightly earlier spectral type.

IGR J16393-4643 was observed in the near-infrared on 2004 July 9 (MJD\,53195.3) using the 3.5\,m New Technology Telescope (NTT) at La Silla Observatory \citep{2008A&A...484..783C}.  While \citet{2008A&A...484..783C} conclude that the spectral type of the donor star is BIV-V based on the spectral features and spectral energy distribution (SED), we note that \citet{2008ATel.1450....1N} proposed the donor star to be a K or M supergiant using the same SED. Furthermore, observations with the \textsl{Chandra} observatory show that the previously proposed counterpart is positionally inconsistent with the X-ray source \citep{2012ApJ...751..113B}.  Using the GLIMPSE survey, \citet{2012ApJ...751..113B} proposed that the counterpart must be a distant reddened B-type star.

\subsubsection{IGR J16418-4532}
\label{IGR J16418-4532 Candidates}

Our results show that while the derived masses and radii for the previously proposed spectral types satisfy the eclipse half-angle (see Table~\ref{J16418 Primary Parameters}), the radius would be larger than the Roche-lobe size \citep{1983ApJ...268..368E}.  An O8 I star has a mass of 28\,$M_\sun$ according to \citet{2006ima..book.....C} and \citet{2000asqu.book.....A}.  We find the maximum radius for a 28\,$M_\sun$ star to satisfy the constraint imposed by the Roche-lobe to be 18.2\,$R_\sun$.  This is clearly smaller than the radii reported in \citet{2006ima..book.....C} and \citet{2000asqu.book.....A}, which are 22\,$R_\sun$ and 20\,$R_\sun$, respectively (see Table~\ref{J16418 Primary Parameters}).  Since the radius for each proposed spectral type is too large to satisfy the constraint imposed by the Roche-lobe, it is our contention that the donor must be an earlier spectral type.

We find that spectral classes O7.5 I and earlier satisfy the constraint imposed by the Roche-lobe radius (see Table~\ref{IGR J16418-4532 Candidate Parameters}).  The ratio between the radius of the donor star and that of the Roche-lobe, $\beta$, is found to exceed 0.9, which is consistent with other HMXBs that host supergiants {\mybf \citep{1984ARA&A..22..537J}}.  Transitional Roche-lobe accretion has been proposed in IGR J16418-4532 where a fraction of the mass transfer is due to a focused wind \citep{2012MNRAS.420..554S}.  A focused wind or accretion stream requires a mass donor that nearly fills the Roche-lobe.  If this is the case, we would expect variations in the mass accretion rate that would be attributable to a focused wind or accretion stream \citep{1997ASPC..121..361B}. This mechanism would lead to large variability in the X-ray intensity and could be observed in folded-light curves as well as hardness-intensity diagrams.  Large intensity swings on the order of $\sim$100 were observed in both \textsl{Swift} and \textsl{XMM-Newton} observations of IGR J16418-4532 \citep{2012MNRAS.420..554S,2013ATel.5131....1D}.

% Table 11
\begin{deluxetable}{cccccccccccc}
\tablecolumns{12}
\tablewidth{0pc}
\tablecaption{Possible Parameters of Candidate Donor Stars for IGR J16418-4532}
\tablehead{
\colhead{Spectral Type} & \colhead{$M/M_\sun$} & \colhead{$q$} & \colhead{$R/R_\sun$} & \colhead{$R_{\rm L}$$/R_\sun$$^a$} & \colhead{$R_{\rm L}$$/R$} & \colhead{$M_{\rm V}$} & \colhead{$(J-K)_{\rm 0}$$^a$} & \colhead{$E(J-K)$$^a$} & \colhead{$i$$\degr$$^b$} & \colhead{$d_{\rm sun}$$^c$} & \colhead{$d_{\rm sun}$$^d$} \\
\colhead{} & \colhead{} & \colhead{} & \colhead{} & \colhead{} & \colhead{} & \colhead{} & \colhead{} & \colhead{} & \colhead{} & \colhead{(kpc)} & \colhead{(kpc)}} 
\startdata
\textsl{O6 Ia} & 44.10 & 0.032 & 19.95 & {\mybf 23.85} & {\mybf 0.84} & \textsl{-6.38} & -0.21 & 2.60 & {\mybf 77--81} & 11.9 & 11.9 \\
\textsl{O6.5 Ia} & 41.20 & 0.034 & 20.22 & {\mybf 23.18} & {\mybf 0.87} & \textsl{-6.38} & -0.21 & 2.60 & {\mybf 73--76} & 12.0 & 12.1 \\
\textsl{O7 Ia} & 38.44 & 0.036 & 20.49 & {\mybf 22.52} & {\mybf 0.91} & \textsl{-6.38} & -0.21 & 2.60 & {\mybf 70--73} & 12.0 & 12.3 \\
\textsl{O7.5 Ia} & 36.00 & 0.039 & 20.79 & {\mybf 21.90} & {\mybf 0.95} & \textsl{-6.38} & -0.21 & 2.60 & {\mybf 67--69} & 12.0 & 12.4 \\
\textsl{O8 Ia} & 33.72 & 0.042 & 21.10 & {\mybf 21.31} & {\mybf 0.99} & \textsl{-6.38} & -0.21 & 2.60 & {\mybf 64--66} & 11.9 & 12.6 \\
\enddata
\tablecomments{Possible Parameters of Candidate Donor Stars. \\*
$^a$ The value for $(J-K)_{\rm 0}$ was calculated using $(J-H)_{\rm 0}$ and $(H-K)_{\rm 0}$ published in \citet{2006A&A...457..637M}. $E(J-K)$ is found by subtracting $(J-K)_{\rm 0}$ from the observed $J-K$ \\*
$^b$ The range of inclination angles of the system consistent with the measured eclipse half-angle. \\*
$^c$ The distance the object is from the Sun using the distance modulus. \\*
$^d$ The distance the object is from the Sun using the radius to distance ratio derived from spectral energy distribution. The radius to distance ratio is found to be 3.77$\times$10$^{-11}$ \citep{2008A&A...484..783C}. \\*
}
\label{IGR J16418-4532 Candidate Parameters}
\end{deluxetable}

%figure 19
\begin{figure}
\centerline{\includegraphics[width=3in]{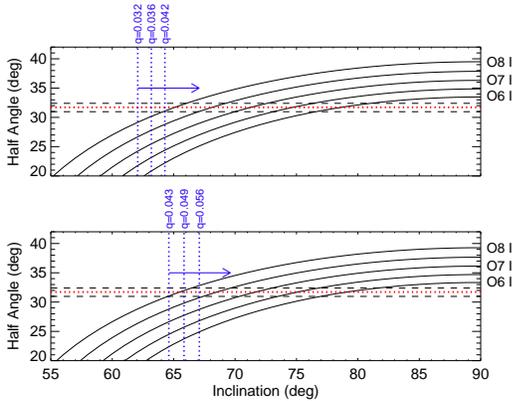}}
\figcaption[march30-2015_IGRJ16418-4532_candidate_eclipse_param.eps]{
The black curves show the predicted eclipse half angle as a function of inclination angle for stars with the candidate spectral types for IGR J16418-4532.  The red and black dashed lines indicate the eclipse half angle and estimated error as measured by \textsl{Swift} BAT.  We assume a neutron star mass of 1.4\,$M_\sun$ (top) and of mass 1.9\,$M_\sun$ (bottom) and typical masses and radii for the assumed companion spectral type (see Table~\ref{IGR J16479-4514 Candidate Parameters}).  The blue vertical dashed lines indicate the lower limit of the inclination angle.  Inclinations to the left of these correspond to stars that overfill the Roche-lobe.
\label{J16418 Candidate Half Angle}
}
\end{figure}

Near-infrared spectral features previously led to a spectral classification of either a late O-type supergiant \citep{2008A&A...484..783C} or a BN0.5 Ia \citep{2013A&A...560A.108C}.  We note the presumed radius of both spectral types overfills the Roche-lobe, which shows that the proposed spectral classifications are incorrect.  The spectral type of the mass donor places an additional constraint on the source distance.  Under the assumption that the K-band magnitude ($m_{\rm K}$) and extinction in the V-band ($A_{\rm V}$) are magnitudes 11.48 and 14.5 for an O8.5 I classification, \citep{2008A&A...484..801R} found the distance of the source to be $\sim$13\,kpc.  Converting $A_{\rm V}$ to $A_{\rm K}$ using Table 3 in \citet{1985ApJ...288..618R}, we confirm the calculations for the distance of IGR J16418-4532 in \citet{2008A&A...484..801R} using the distance modulus.  The distance of IGR J16418-4532 assuming the aforementioned spectral types are reported in Table~\ref{IGR J16418-4532 Candidate Parameters} using the values for $M_{\rm V}$ obtained from \citet{2006MNRAS.371..185W} and \citet{2006A&A...457..637M}.  We also use the radius to distance ratio from the near-infrared spectral energy distribution (SED) measurements reported in \citep{2008A&A...484..801R} together with our eclipse measurements to estimate the distance of IGR J16418-4532.  We find the distance to be between 11.9--12.6\,kpc, which is slightly smaller than that reported in \citet{2008A&A...484..801R} for an O8 I star.

IGR J16418-4532 is a heavily absorbed SFXT where the observed $N_{\rm H}$, measured to be between 3.9$^{+1.2}_{-0.9}$$\times$10$^{22}$\,atoms cm$^{-2}$ and 7$\pm$2$\times$10$^{22}$\,atoms cm$^{-2}$ \citep{2012MNRAS.419.2695R}, exceeds the values reported by the Leiden/Argentine/Bonn survey \citep{2005A&A...440..775K} and in the review by \citet{1990ARA&A..28..215D}, which are 1.59$\times$10$^{22}$ and 1.88$\times$10$^{22}$\,atoms cm$^{-2}$, respectively.  Since the measured $N_{\rm H}$ was found to be in excess of the Galactic H I, determining the interstellar fraction of $N_{\rm H}$ is problematic and the value in \citet{2008A&A...484..801R} for the extinction cannot be verified without the presence of systematic error.

We compare the observed value of the $J-K$ color of 2.39\footnote{\url{http://www.iasfbo.inaf.it/\~masetti/IGR/sources/16418.html}} with the intrinsic $(J-K)_{\rm 0}$ for the proposed mass donors (see Table~\ref{IGR J16418-4532 Candidate Parameters}).  Calculating the difference between the observed $J-K$ and the intrinsic $(J-K)_{\rm 0}$, we find the reddening values $E(J-K)$ for each proposed spectral type for the mass donor (see Table~\ref{IGR J16418-4532 Candidate Parameters}).  We calculate the reddening in the $B-V$ band, $E(B-V)$, using Equation 1 in \citet{2009MNRAS.400.2050G} and the extinction in the V-band ($A_{\rm V}$) \citep{2008A&A...484..801R}.  Converting $E(B-V)$ to $E(J-K)$ using Table 3 in \citet{1985ApJ...288..618R}, we find $E(J-K)$ to be 2.45.  While this is slightly lower than what would be expected for late O supergiant spectral types (see Table~\ref{IGR J16418-4532 Candidate Parameters}), the presence of systematic error described above prevents an exact calculation.

These results show that stars with spectral type O7.5 I and earlier satisfy the constraints imposed by both the duration of the eclipse and the Roche-lobe (see Figure~\ref{J16418 Candidate Half Angle}).  Since the measured $N_{\rm H}$ is largely in excess of interstellar values measured by \citet{2005A&A...440..775K} and \citet{1990ARA&A..28..215D}, determining what fraction of $N_{\rm H}$ is interstellar in origin is problematic.  We find measurements such as the distance to be consistent with the distances determined by \citet{2008A&A...484..801R}.  Given our measurements, spectral types near O7.5 I are reasonable.  By constraining this type of star, we solved the first part of a three part problem.  The remaining pieces include pulse-timing measurements which can be done in the X-ray and radial velocity measurements in the near-infrared.

\subsubsection{IGR J16479-4514}

Our results show that while the expected masses and radii for the previously proposed spectral types for IGR J16479-4514 satisfy the eclipse half-angle (see Table~\ref{J16479 Primary Parameters}), the implied radius would be larger than the Roche-lobe radius \citep{1983ApJ...268..368E}.  This is similar to the situation observed in IGR J16418-4532, which we describe in Section~\ref{IGR J16418-4532 Candidates}.  We calculate the Roche-lobe radius for stars of spectral types O8.5 I and O9.5 Iab to be 19.1\,$R_\sun$ and 18.1\,$R_\sun$, respectively (see Table~\ref{J16479 Primary Parameters}).  Since the radius for each proposed spectral type is too large to satisfy the constraint imposed by the Roche-lobe (see Table~\ref{J16479 Primary Parameters}), we suggest that the donor must be a slightly earlier spectral type.

We find that spectral classes O7 I and earlier satisfy the Roche-lobe constraint (see Table~\ref{IGR J16479-4514 Candidate Parameters}).  The ratio between the radius of the donor star and that of the Roche-lobe, $\beta$, is found to exceed 0.9, which is consistent with other HMXBs that host supergiants {\mybf \citep{1984ARA&A..22..537J}}.  Transitional Roche-lobe has been proposed in IGR J16479-4514 \citep{2013MNRAS.429.2763S}, which requires a mass donor that nearly fills the Roche-lobe.  \citet{2013MNRAS.429.2763S} found phase-locked flares were found in their observation of IGR J16479-4514. \citet{2013MNRAS.429.2763S} attributes the physical mechanism responsible for these flares which are spaced 0.2 in phase as large-scale structures in the wind.

% Table 12
\begin{deluxetable}{cccccccccccc}
\tablecolumns{12}
\tablewidth{0pc}
\tablecaption{Possible Parameters of Candidate Donor Stars for IGR J16479-4514}
\tablehead{
\colhead{Spectral Type} & \colhead{$M/M_\sun$} & \colhead{$q$} & \colhead{$R/R_\sun$} & \colhead{$R_{\rm L}$$/R_\sun$$^a$} & \colhead{$R_{\rm L}$$/R$} & \colhead{$M_{\rm V}$} & \colhead{$(J-K)_{\rm 0}$$^a$} & \colhead{$E(J-K)$$^a$} & \colhead{$i$$\degr$$^b$} & \colhead{$d_{\rm sun}$$^c$} & \colhead{$d_{\rm sun}$$^d$} \\
\colhead{} & \colhead{} & \colhead{} & \colhead{} & \colhead{} & \colhead{} & \colhead{} & \colhead{} & \colhead{} & \colhead{} & \colhead{(kpc)} & \colhead{(kpc)}} 
\startdata
\textsl{O6 Ia} & 44.10 & 0.032 & 19.95 & {\mybf 22.03} & 0.91 & \textsl{-6.38} & -0.21 & 3.48 & {\mybf 69--75} & 4.44 & 4.50 \\
\textsl{O6.5 Ia} & 41.20 & 0.034 & 20.22 & 21.42 & 0.94 & \textsl{-6.38} & -0.21 & 3.48 & {\mybf 66--71} & 4.46 & 4.56 \\
\textsl{O7 Ia} & 38.44 & 0.036 & 20.49 & {\mybf 20.80} & {\mybf 0.99} & \textsl{-6.38} & -0.21 & 3.48 & {\mybf 63--68} & 4.46 & 4.62 \\
\textsl{O7.5 Ia} & 36.00 & 0.039 & 20.79 & {\mybf 20.23} & {\mybf 1.03} & \textsl{-6.38} & -0.21 & 3.48 & {\mybf 60--64} & 4.48 & 4.69 \\
\textsl{O8 Ia} & 33.72 & 0.042 & 21.10 & {\mybf 19.68} & {\mybf 1.07} & \textsl{-6.36} & -0.21 & 3.48 & {\mybf 58--61} & 4.44 & 4.81 \\
\enddata
\tablecomments{Possible Parameters of Candidate Donor Stars. \\*
$^a$ The value for $(J-K)_{\rm 0}$ was calculated using $(J-H)_{\rm 0}$ and $(H-K)_{\rm 0}$ published in \citet{2006A&A...457..637M}. $E(J-K)$ is found by subtracting $(J-K)_{\rm 0}$ from the observed $J-K$ \\*
$^b$ The range of inclination angles of the system consistent with the measured eclipse half-angle. \\*
$^c$ The distance the object is from the Sun using the distance modulus. \\*
$^d$ The distance the object is from the Sun using the radius to distance ratio derived from spectral energy distribution. The radius to distance ratio is found to be 1$\times$10$^{-10}$ \citep{2008A&A...484..783C}. \\*
}
\label{IGR J16479-4514 Candidate Parameters}
\end{deluxetable}

%figure 20
\begin{figure}[ht]
\centerline{\includegraphics[width=3in]{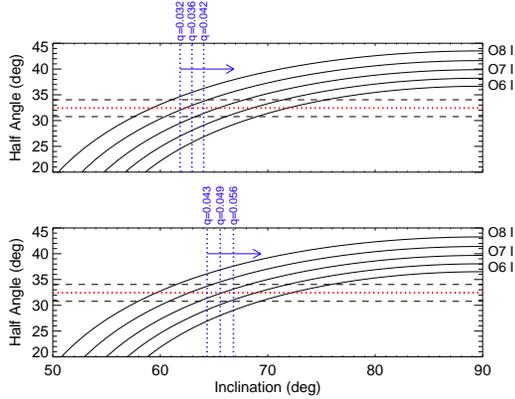}}
\figcaption[april29-2015_IGR_J16479-4514_candidate_eclipse_param.eps]{
The black curves show the predicted eclipse half angle as a function of inclination angle for stars with the candidate spectral types for IGR J16479-4514.  The red and black dashed lines indicate the eclipse half angle and estimated error as measured by \textsl{Swift} BAT.  We assume a neutron star mass of 1.4\,$M_\sun$ (top) and of mass 1.9\,$M_\sun$ (bottom) and typical masses and radii for the assumed companion spectral type (see Table~\ref{IGR J16479-4514 Candidate Parameters}).  The blue vertical dashed lines indicate the lower limit of the inclination angle.  Inclinations to the left of these correspond to stars that overfill the Roche-lobe.
\label{J16479 Candidate Half Angle}
}
\end{figure}

Using near-infrared spectral features, \citet{2008A&A...484..783C} previously determined the spectral classification of the donor star to be a late O-type supergiant \citep{2008A&A...484..783C}.  The donor spectral type places an additional constraint on the distance of IGR J16479-4514.  The K-band magnitude ($m_{\rm K}$) and the extinction in the V-band ($A_{\rm V}$) are found to be 9.79 and 18.5 \citep{2008A&A...484..801R}.  Additionally, the $R_*/D_*$ ratio was found to be 1$\times$10$^{-10}$ \citep{2008A&A...484..783C}.  Using this information, the minimum distance of IGR J16479-4514 was found to be $\sim$4.9\,kpc \citep{2008A&A...484..783C}.  Converting $A_{\rm V}$ to $A_{\rm K}$ using Table 3 in \citet{1985ApJ...288..618R}, we find $A_{\rm K}$ to be 2.07 and confirm the calculation for the distance of IGR J16479-4514 using the distance modulus.  The distances of IGR J16479-4514 assuming the aforementioned spectral types are reported in Table~\ref{IGR J16479-4514 Candidate Parameters} using the values for $M_{\rm V}$ obtained from \citet{2006MNRAS.371..185W} and \citet{2006A&A...457..637M}.  We find that the distance of stars with spectral type O7 I and earlier to be between 4.4--4.6\,kpc, which are resonably consistent with the measurements in \citet{2008A&A...484..801R}.

IGR J16479-4514 is a heavily absorbed SFXT where the $N_{\rm H}$ was measured to be 9.5$\pm$0.3$\times$10$^{22}$\,atoms cm$^{-2}$ \citep{2013MNRAS.429.2763S}.  This is an order of magnitude larger than the Galactic H I values reported by the Leiden/Argentine/Bonn survey \citep{2005A&A...440..775K} and in the review by \citet{1990ARA&A..28..215D}, which are 1.87$\times$10$^{22}$ and 2.14$\times$10$^{22}$\,atoms cm$^{-2}$, respectively.  Since the measured $N_{\rm H}$ for both sources was found to be in excess of the Galactic H I, determining the interstellar fraction of $N_{\rm H}$ is difficult.  Therefore, we cannot verify the value for the extinction in \citet{2008A&A...484..801R} without the presence of systematic error.

We calculate the reddening values $E(J-K)$ for each proposed spectral type for the mass donor (see Table~\ref{IGR J16479-4514 Candidate Parameters}) using the observed value of the $J-K$ color of 3.27\footnote{\url{http://www.iasfbo.inaf.it/\~masetti/IGR/sources/16479.html}} and the intrinsic $(J-K)_{\rm 0}$ for the proposed mass donors (see Table~\ref{IGR J16479-4514 Candidate Parameters}).  To check for consistency with late O supergiant stars, we compare this with the reddening in the $B-V$ band using Equation 1 in \citet{2009MNRAS.400.2050G} and the extinction in the V-band ($A_{\rm V}$) \citep{2008A&A...484..801R}. Converting $E(B-V)$ to $E(J-K)$ using Table 3 in \citet{1985ApJ...288..618R}, we find the reddening in the $E(J-K)$ band to be 3.12.  While this is slightly lower than what would be expected for late O supergiant spectral types, the presence of systematic errors, described above prevents an exact calculation.

Stars with spectral type O7 I and earlier satisfy the eclipse duration and Roche-lobe constraints (see Figure~\ref{J16479 Candidate Half Angle}).  Determining the interstellar fraction of $N_{\rm H}$ is difficult since the measured $N_{\rm H}$ is greatly in excess of interstellar values.  We find the distance of the proposed counterparts to be consistent with those determined by \citet{2008A&A...484..801R}.  Since no pulsation period was identified, the next step to constrain the donor star would be radial velocity measurements in the near-infrared.

\subsubsection{IGR J18027-2016}

Our results show that the mass and radius of the donor star in IGR J18027-2016 can be constrained to be between {\mybf 18.6$\pm$0.9}\,$M_\sun$ and {\mybf 17.4$\pm$0.9}\,$R_\sun$ and {\mybf 19.4$\pm$0.9}\,$M_\sun$ and {\mybf 19.5$^{+0.8}_{-0.7}$}\,$R_\sun$ for edge-on orbits and inclinations where the Roche-lobe is completely filled, respectively.  We find the inclination where the Roche-lobe is filled to be {\mybf 73.3}$\degr$, which is consistent with the earlier results from \citet{2011A&A...532A.124M}.  Since the {\mybf semi-amplitude of the} radial velocities of both the donor star and compact object are known, we also find that the mass of the neutron star can be constrained.  We calculate the mass of the neutron star to be between {\mybf 1.37$\pm$0.19}\,$M_\sun$ and {\mybf 1.43$\pm$0.20}\,$M_\sun$ for our lower and upper limits (see Section~\ref{IGR J18027-2016 Results}).

Since the radius of the donor star is constrained, we also can estimate the distance, optical and near infrared magnitudes of IGR J18027-2016.  Using SED measurements, \citet{2008A&A...484..783C} calculated the radius to distance ratio to be 4$\pm$1$\times$10$^{-11}$, where the uncertainties are at the 90$\%$ confidence interval.  At the 90$\%$ confidence interval, we find the eclipse half-angle to be {\mybf 34$_{-3}^{+4}$$\degr$} and the radius of the donor star to be constrained to between {\mybf 17$^{+2}_{-1}$\,$R_\sun$} and {\mybf 19$\pm$1\,$R_\sun$} in the two limits described in Section~\ref{IGR J18027-2016 Results}.  Combining our results for the radius of the donor star with the radius to distance ratio \citep[Table 6;][]{2008A&A...484..783C}, we find that the distance of IGR J18027-2016 can be constrained to {\mybf 11$\pm$2\,kpc} and {\mybf 12$\pm$2\,kpc} in the two limits.  Using the distance modulus \citep[e.g.][]{2008A&A...486..911N}, the absolute magnitude of IGR J18027-2016 can be calculated.  The apparent magnitude in the R-band, $R$, and the extinction in the V-band, $A_{\rm V}$ were found to be 16.9 and 8.5 \citep{2008A&A...482..113M}.  We find the absolute magnitude in the R-band, $M_{\rm R}$ to be $\sim$-5 at both constraints, which is what would be expected for a B-type supergiant \citep{2006A&A...457..637M,2006MNRAS.371..185W}.

Our results are consistent with the previously proposed B1 Ib \citep{2010A&A...510A..61T} or B0-B1 I \citep{2011A&A...532A.124M} spectral types in IGR J18027-2016.  We constrain the mass and radius of the donor star to be between {\mybf 18.6--19.4}\,$M_\sun$ and {\mybf 17.4--19.5}\,$R_\sun$.  We also constrain the mass of the neutron star to be {\mybf 1.18--1.63}\,$M_\sun$, which {\mybf marginally} improves on the results in \citet{2011A&A...532A.124M}{\mybf , which was found to be 1.15--1.85\,$M_\sun$}.

\subsubsection{XTE J1855-026}

{\mybf We find that the mass and radius of the donor star in XTE J1855-026 are constrained between 19.6$\pm$1.1\,$M_\sun$ and 21.5$\pm$0.5\,$R_\sun$ at edge-on orbits to 20.2$\pm$1.2\,$M_\sun$ and 23.0$\pm$0.5\,$R_\sun$ where the Roche-lobe is just filled.  The inclination where the Roche-lobe is filled is found to be 76.4$\degr$.  We find the derived masses and radii to be consistent with those reported in \citet{2006ima..book.....C} and \citet{2000asqu.book.....A} for stars with spectral type B0 I.  Since the {\mybf semi-amplitude of the} radial velocities for both the donor star and compact object are known, we find that the mass of the neutron star can be constrained to be between 1.77$\pm$0.55\,$M_\sun$ and 1.82$\pm$0.57\,$M_\sun$ (see Section~\ref{XTE J1855-026 Results}).}  {\mybf We note that the large error estimates in the mass of the neutron star are likely to be attributed to substantial uncertainties in the estimate in $K_{\rm o}$ as reported in \citet{2015arXiv150301087G}.  This is likely to be attributed to emission line contamination and/or changes in the stellar wind.}

Based on optical and near-infrared spectra together with the analysis reported in \citet{2002ATel..102....1V}, the spectral type of the donor star was previously determined to be a B0 Iaep \citep{2008ATel.1876....1N}. {\mybf Based on the ratio of the equivalent widths of Si IV to Si III which is a diagnostic for supergiant spectral types \citep{1971ApJS...23..257W}, the mass donor was recently refined to a BN0.2 Ia by \citet{2015arXiv150301087G}.} Using SED measurements, \citet{2013ApJ...764..185C} calculated the radius and distance to be 26.9\,$R_\sun$ and 10.8$\pm$1.0\,kpc, which places XTE J1855-026 in the Scutum arm region.  {\mybf Combining the properties of the newly derived spectral type with the distance modulus, \citet{2015arXiv150301087G} recently calculated the nominal distance of XTE J1855-026 to be 10$^{+7}_{-4}$\,kpc.}  Based on these results, the radius to distance ratio can be calculated to be 5.6$\pm$0.5$\times$10$^{-11}$.  Combining our results for the radius of the donor star with the calculated radius to distance ratio, we find the distance of XTE J1855-026 can be constrained to {\mybf 8.6$\pm$0.8\,kpc} and {\mybf 9.2$\pm$0.9\,kpc} in the two limits.

\citet{1999ApJ...517..956C} measured the $N_{\rm H}$ in XTE J1855-026 to be 14.7$\pm$0.6$\times$10$^{22}$\,atoms cm$^{-2}$, which exceeds the values reported by the Leiden/Argentine/Bonn survey \citep{2005A&A...440..775K} and in the review by \citet{1990ARA&A..28..215D}.  In a \textsl{Swift} XRT observation, the $N_{\rm H}$ was found to be 4.1$\pm$0.5$\times$10$^{22}$\,atoms cm$^{-2}$ \citep{2008ATel.1875....1R}.  These are signifcantly larger than the measurements for the interstellar $N_{\rm H}$, which are 6.62$\times$10$^{21}$\,atoms cm$^{-2}$ \citep{2005A&A...440..775K} and 7.35$\times$10$^{21}$\,atoms cm$^{-2}$ \citep{1990ARA&A..28..215D}.  Since the measured $N_{\rm H}$ was found to be in excess of the Galactic H I, the conversion between the interstellar $N_{\rm H}$ and the extinction $A_{\rm V}$ is problematic and the value of 5.8$\pm$0.9 in \citet{2013ApJ...764..185C} cannot be verified without the presence of systematic error.

\subsection{What is the nature of the non-zero eclipse flux in IGR J16393-4643?}
\label{What is the nature of the non-zero eclipse flux in IGR J16393-4643?}

The source flux during eclipse in IGR J16393-4643 does not reach 0\,counts s$^{-1}$ in the folded light curves (see Figure~\ref{J16393 Folded Half Angle}).  We find the ratio between the flux during eclipse to the flux outside eclipse to be {\mybf 54$\pm$5$\%$}. This is significantly larger than what is observed in the other XRBs in our study (see Table~\ref{Step and Ramp Model}).  

We first discuss the possible scenario where we observe a partial eclipse in IGR J16393-4643.  In our model for the eclipse (see Section~\ref{Eclipse Modeling}), we assume the compact object to be a point source {\mybf \citep{1984ARA&A..22..537J}}.  While this is a valid assumption since the radius of the compact object is much smaller than that of the donor star, our approximation does not consider the extended X-ray emission region.  In this case, the constraints the mass and radius of the donor star must take into account the size of the extended emission region.  Since a significant residual flux is observed, it is likely that we observe a partial eclipse.

We also consider the possibility that the residual emission observed in IGR J16393-4643 is attributed to a dust-scattering halo similar to what is observed in some other HMXBs \citep[Cen X-3; Vela X-1; OAO 1657-415,][]{1991MNRAS.251...76D,1994ApJ...436L...5W,2006MNRAS.367.1147A}.  While residual emisison has been attributed to a dust-scattering in some other HMXBs, the residual emission found in IGR J16393-4643 in the BAT folded light curves (15--50\,keV) is seen at much higher levels (see Table~\ref{Step and Ramp Model}).  A dust-scattering halo is predominantly a soft X-ray phenomenon.  While a significant fraction of the out-of-eclipse flux may be in a dust-scattering halo \citep{1995A&A...293..889P}, we expect a smaller fraction to be in the energies resolved by BAT \citep[e.g.][and references therein]{2014ApJ...793...77C}.  Therefore, we conclude a dust-scattering halo cannot be the sole mechanism responsible for the residual emission seen in the folded light curves.

Finally, we discuss the possibility that Compton scattering and reprocessing in a region of dense gas could account for the residual flux in eclipse.  \citet{2006A&A...447.1027B} found that a Compton emission (\texttt{comptt} within Xspec) model with an electron temperature of 4.4$\pm$0.3\,keV and optical depth of 9$\pm$1 provides a good fit to the average spectrum of IGR J16393-4643.  Additionally, Fe K$\alpha$ and Fe K$\beta$ lines were found at 6.4\,keV and 7.1\,keV, respectively, where the ratio of the iron intensities is seen to be consistent with photoionization \citep{2006A&A...447.1027B,1993A&AS...97..443K,2015MNRAS.446.4148I}.  The equivalent widths (EQW) {\mybf in the \textsl{XMM-Newton} observation} were found to be 60$\pm$30\,eV and an upper limit of 120\,eV, for the Fe K$\alpha$ and Fe K$\beta$ lines respectively \citep{2006A&A...447.1027B}.  {\mybf In a recent \textsl{Suzaku} observation, the EQW of the Fe K$\alpha$ and Fe K$\beta$ lines were found to be 46$^{+7}_{-6}$\,eV and an upper limit of 30\,eV \citep{2015MNRAS.446.4148I}}.  While the mechanism responsible for the Fe K$\alpha$ and Fe K$\beta$ is likely to be fluorescence of cold matter, the equivalent widths point to a likely origin in a spherical distribution of dense gas \citep{2006A&A...447.1027B}.  Therefore, Compton scattering and reprocessing might be the sole contributor to the residual emission observed in the BAT folded light curves.

It is likely that a combination of the mechanisms described above account for the residual emission found in the folded light curves of IGR J16393-4643 (see Section~\ref{IGR J16393-4643 Results}).  Since the count rate during eclipse is significantly larger than what is observed in most other eclipsing HMXBs, it is likely that only a small fraction comes from a dust-scattering halo.

\subsection{What mechanism is responsible for asymmetries in the eclipse profile?}
\label{Constraints on eccentricity}

{\mybf The eclipse profiles show the presence of asymmetries (see Tables~\ref{Asymmetric Step and Ramp Model} and~\ref{Historic Step and Ramp Model}) as previously noted by \citet{2015A&A...577A.130F} in the cases of IGR J18027-2016 and XTE J1855-026 and \citet{2009MNRAS.397L..11J} in the case of IGR J16479-4514.  These asymmetries seen in the ingress and egress durations are suggestive of the presence of complex structures in the wind such as accretion or photoionization wakes.}
 
We first discuss the possible case that the asymmetry in the eclipse profiles can be attributed to accretion wakes.  In an HMXB driven entirely by a spherical wind, material is only accreted in a cylindrical region where the kinetic energy is less than the gravitational potential energy of the compact object.  The radius of the accretion cylinder is the Bondi-Hoyle accretion radius  \citep[Equation 1,][]{1996A&A...311..793F}.  Perturbed material forms an ``accretion wake" that typically trails the orbit of the neutron star and results in large intrinsic column densities \citep{1990ApJ...356..591B}.  Prior to eclipse, the progressively increasing $N_{\rm H}$ partially obscures the X-ray emission resulting in longer ingress durations compared to the duration of egress.  Since the accretion wake is located beyond the compact object during egress, no apparent increase in the intrinsic $N_{\rm H}$ is observed.  The ingresses observed are somewhat larger than egress, which is consistent with the presence of an accretion wake \citep{1990ApJ...356..591B}.  The count rate prior to ingress is also somewhat smaller than that observed after egress, providing additional evidence for accretion wakes.  Hardness ratios or measurements of $N_{\rm H}$ folded on the orbital period could be implemented to confirm the possibility of accretion wakes.

We also consider the possibility that photoionization wakes could explain the asymmetric eclipse profiles.  The eclipse profiles of IGR J18027-2016 and XTE J1855-026 are compared to those seen in eclipsing systems where asymmetric density enhancements are observed on large spatial dimensions \citep[e.g. Vela X-1;][]{1996A&A...311..793F}.  In systems where the X-ray luminosity is significantly high, a switch-off of the radiative driving force could lead to a reduced wind velocity and enhanced wind density \citep{1996A&A...311..793F}.  This enhanced X-ray scattering region trails the neutron star and results in ingress durations that are significantly larger than those observed at egress \citep{1996A&A...311..793F}.  The eclipse profiles of IGR J18027-2016 and XTE J1855-026 differ from those expected from a photoionization wake--the ingress duration in Vela X-1 was seen to be $\phi=$0.11 \citep{1996A&A...311..793F}.

{\mybf We additionally discuss how energy dependence in the asymmetric eclipse profiles can arise.  The high $N_{\rm H}$, on the order of 10$^{23}$\,atoms cm$^{-2}$, implies the presence of a strong circumstellar wind \citep{2005AIPC..797..402K}, and the X-ray absorption due to this will lead to sharper ingress and egress transitions compared to that seen at lower energies.  In a \textsl{Suzaku} observation of IGR J16479-4514 that covers part of the eclipse, \citet{2013MNRAS.429.2763S} found the egress transition to be broader in soft energies.  In their review, \citet{2015A&A...577A.130F} found asymmetries to be slightly enhanced at lower energies compared to higher energies.  While investigating the energy dependence in the eclipse transitions is beyond the scope of the present paper, this study in obscured SGXBs will be difficult due to a reduced count rate at low energies that results from the large intrinsic absorption present in these systems \citep{2015A&A...577A.130F}.}

\begin{figure}[ht]
\centerline{\includegraphics[width=3in]{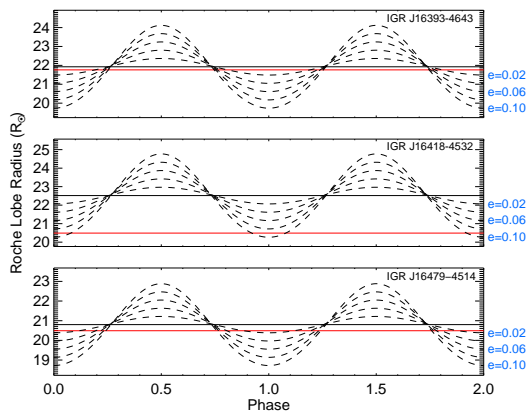}}
\figcaption[may4-2015_No-RV_eccentric-test.eps]{
A sample plot of the Roche-lobe of the donor stars in IGR J16393-4643 (top), IGR J16418-4532 (middle) and IGR J16479-4514 (bottom) vs. orbital phase for eccentricities ranging from 0.0 to 0.1 in steps of 0.02.  The horizontal red line represents the radius of the donor star under the assumption that the donor is an O9 I for IGR J16393-4643 and O7 I for both IGR J16418-4532 and IGR J16479-4532 \citep{2005A&A...436.1049M}.
\label{Non Radial-Velocitiy Eccentricity Constraint}
}
\end{figure}

{\mybf Finally, we consider the possibility that the asymmetric eclipse profiles could be attributed to a small to modest eccentricity.  Since we considered objects with relatively short orbital periods, we expect the eccentricity of the systems to be near zero \citep{1977A&A....57..383Z,2014MNRAS.440.1626M}.  The eccentricity in IGR J18027-2016 and XTE J1855-026 were both noted to be small to modest, where $e$ was found to be less than 0.2 and 0.04, respectively \citep{2003ApJ...596L..63A,2002ApJ...577..923C}.  In the cases of IGR J16393-4643, IGR J16418-4532 and IGR J16479-4514 where no pulse-timing or radial-velocity methods are available to determine an orbital solution, we constrain the maximum allowed eccentricity to that where the radius of the donor star completely fills the Roche-lobe at periastron \citep[][and references therein]{2013MNRAS.434.2182G}.  These were all found to be near zero (see Figure~\ref{Non Radial-Velocitiy Eccentricity Constraint}).  Since the eccentricities are found to be small to modest, we do not expect these to result in sizeable asymmetries in the eclipse profile.  Additionally, apsidal advance will be apparent in the case that an eccentric orbit could lead to asymmetries in the eclipse profile.  While apsidal advance would be difficult to detect in the $\sim$9\,yr of \textsl{Swift} data, we believe it to be unlikely for asymmetries to be solely attributed to small to modest eccentricities.  Furthermore, accurate measurements of apsidal advance will depend on multiple pulse-timing measurements of these systems, which are not yet available.}

%figure 21

\section{Conclusion}

Eclipsing X-ray binaries provide an opportunity to constrain the physical parameters of the donor star as well as the compact object.  To determine the eclipse half-angle in our survey, we modeled the eclipses using {\mybf both} symmetric {\mybf and asymmetric} step-and-ramp {\mybf functions}.  The luminosity of each system is less than expected for Roche-lobe overflow \citep{2004RMxAC..21..128K}, which means we can attach the constraint that the mass donor underfills the Roche-lobe.  Since IGR J18027-2016 {\mybf and XTE J1855-026 are the only ``double-lined binaries"} in our sample, we calculate the parameters of the other systems assuming the neutron stars to be at the white-dwarf Chandrasekhar limit, 1.4$M_\sun$.  We also calculated the parameters of the other systems assuming a more massive neutron star--1.9$M_\sun$.

Our results show that stars with spectral type B III satisfy both constraints imposed by the eclipse duration and Roche-lobe for IGR J16393-4643.  Assuming the estimates for the mass and radius of a B5 III star reported in \citet{2006ima..book.....C}, we find the mass and radius of the donor star to exceed 7\,$M_\sun$ and 6.3\,$R_\sun$. B I stars were found to overfill the Roche-lobe.  The source emission in IGR J16393-4643 does not reach 0\,counts s$^{-1}$ in the folded light curves, where the fraction between the flux in eclipse to that outside eclipse was found to be {\mybf 54$\pm$5$\%$} (see Tables~\ref{Step and Ramp Model}--~\ref{Historic Step and Ramp Model}).  Compton scattering and reprocessing in a dense region of gas could possibly account for the X-ray emission region not obscured by the donor star.

Our results show that the previously proposed O8.5 I and BN0.5 Ia spectral types for the mass donor in IGR J16418-4532 must be excluded.  While these spectral types satisfy the eclipse half-angle, the Roche-lobe is significantly overfilled.  Stars with spectral type O7.5 I or earlier are consistent with both the eclipse half-angle and the Roche-lobe.  In this case, we find the mass and radius of the donor star to exceed 36.00\,$M_\sun$ and 20.79\,$R_\sun$ assuming the estimates for the mass and radius of an O7.5 I star reported in \citet{2005A&A...436.1049M}.  We find the minimum inclination angle of the system to be {\mybf 67}$\degr$.  The distance measurements of IGR J16418-4532 are consistent with the previously determined distance (see Table~\ref{IGR J16418-4532 Candidate Parameters}); however, determining the interstellar fraction of $N_{\rm H}$ was found to be problematic.  {\mybf The ingress and egress durations in the folded light curves were found to be asymmetric where the ingress duration is longer than the egress duration.  This is likely attributed to the presence of an accretion wake or from a focused stream as noted in \citet{2012MNRAS.420..554S} and \citet{2013ATel.5131....1D}.}

The previously proposed O8.5 Ia and O9.5 Iab spectral classifications of the mass donor in IGR J16479-4514 must be excluded because the Roche-lobe is significantly overfilled.  However, we found that stars with spectral type O7 I and earlier satisfy both constraints imposed by the eclipse duration and Roche-lobe.  Assuming the estimates for the mass and radius of an O7 I star reported in \citet{2005A&A...436.1049M}, the mass and radius of the donor star are found to exceed 38.44\,$M_\sun$ and 20.49\,$R_\sun$, respectively.  We find the minimum inclination angle of the system to be {\mybf 63$\degr$}.  The distance measurements remain unchanged from earlier measurements (see Table~\ref{IGR J16479-4514 Candidate Parameters}); however, the interstellar fraction of $N_{\rm H}$ remains undetermined.  We find {\mybf the ingress and egress durations to be symmetric within the error bars.}  The ratio between the radius of the donor star and Roche-lobe was found to exceed 0.9, which shows the possibility of transitional Roche-lobe overflow.

The mass and radius of the donor star in IGR J18027-2016 was constrained to {\mybf 18.6$\pm$0.9}\,$M_\sun$ and {\mybf 17.4$\pm$0.9}\,$R_\sun$ and {\mybf 19.4$\pm$0.9}\,$M_\sun$ and {\mybf 19.5$^{+0.8}_{-0.7}$}\,$R_\sun$ in the two limits.  We find the inclination angle where the donor star just fills the Roche-lobe size to be {\mybf 73.3}$\degr$.  We also find the distance measurements of IGR J18027-2016 to be {\mybf 11$\pm$2\,kpc} and {\mybf 12$\pm$2\,kpc} in the allowed limits.  In the allowed limits, we constrained mass of the neutron star to between {\mybf 1.37$\pm$0.19}\,$M_\sun$ and {\mybf 1.43$\pm$0.20}\,$M_\sun$.  The folded light curve shows complicated and asymmetric ingress and egress durations, which can be explained either by the presence of accretion wakes.

Our results show the mass and radius of the donor star in XTE J1855-026 to be constrained to {\mybf 19.6$\pm$1.1\,$M_\sun$ and 21.5$\pm$0.5\,$R_\sun$} at edge-on orbits to {\mybf 20.2$\pm$1.2\,$M_\sun$ and 23.0$\pm$0.5\,$R_\sun$} where the Roche-lobe size is just filled.  We find the inclination angle where the donor star just fills the Roche-lobe size to be {\mybf 76.4\,degrees}.  In the allowed limits, we find the distance of XTE J1855-026 can be constrained to {\mybf 8.6$\pm$0.8}\,kpc and {\mybf 9.2$\pm$0.9}\,kpc.  {\mybf We find the mass of the neutron to be constrained between 1.77$\pm$0.55\,$M_\sun$ and 1.82$\pm$0.57\,$M_\sun$.}  Complicated and asymmetric ingress and egress durations were seen in the folded light curve, which suggests {\mybf the presence of complex structure in the wind}.

To further constrain the physical parameters of the donor star and the compact object in all these systems, additional observations are required.  Constraining the mass of the neutron star will help constrain the Equation-of-State.  Since the pulse period has been accurately measured, for IGR J16393-4643 and IGR J16418-4532 the study would benefit from both pulse-timing analysis as well as radial-velocity curves in the near-infrared.  A radial-velocity curve in the optical or near-infrared would provide additional constraints for IGR J16479-4514 where a pulse-period has yet to be identified.

\acknowledgements

{\mybf We thank the anonymous referee for useful comments and support from NASA 14-ADAP14-0167.}

\end{document}